\documentclass[%
 reprint,
superscriptaddress,
 amsmath,amssymb,
 aps,
]{revtex4-2}

\usepackage{graphicx}% Include figure files
\usepackage{bm}% bold math
\usepackage{mathtools} %to use coloneqq
\usepackage{hyperref}% add hypertext capabilities
\usepackage{xcolor}
\usepackage{booktabs}
\usepackage{xr}
\usepackage{enumitem}
\usepackage{booktabs}
\usepackage{xcolor}
\usepackage{amssymb}
\usepackage{algorithm}
\usepackage{algpseudocode}
\usepackage{ragged2e}
\usepackage{multirow}

\usepackage{amsthm}

\theoremstyle{definition}

\newlength{\savedrulewidth}

\makeatletter
\newcommand*{\addFileDependency}[1]{
  \typeout{(#1)}
  \@addtofilelist{#1}
  \IfFileExists{#1}{}{\typeout{No file #1.}}}
\makeatother

\usepackage{enumerate}
\newcommand{\R}{\mathbb{R}}		% real numbers
\newcommand{\C}{\mathbb{C}}		% complex numbers

\DeclareMathOperator*{\argmin}{arg\,min}

\newcommand{\rev}[1]{\textcolor{black}{#1}}

\begin{document}

\preprint{APS/123-QED}

\title{Generalized Master Stability of Heterogeneous Delay-Coupled Networks}
%Master stability function for heterogeneous delay-coupled networks

\author{Ana Elisa D. Barioni} 
\author{Arthur N. Montanari}
\affiliation{Department of Physics and Astronomy, Northwestern University, Evanston, IL 60208}
\affiliation{Center for Network Dynamics, Northwestern University, Evanston, IL 60208}

\author{Adilson E. Motter}
\affiliation{Department of Physics and Astronomy, Northwestern University, Evanston, IL 60208}
\affiliation{Center for Network Dynamics, Northwestern University, Evanston, IL 60208}
\affiliation{Department of Engineering Sciences and Applied Mathematics, Northwestern University, Evanston, IL 60208}
\affiliation{Northwestern Institute on Complex Systems, Northwestern University, Evanston, IL 60208}

% \date{\today}

\begin{abstract}
Time delays are ubiquitous in physical and biological networked systems, playing a fundamental role in the emergence and stability of collective behavior. Yet, existing theoretical methods for synchronization analysis of delay-coupled systems are largely limited to networks with homogeneous degree distributions. Here, we extend the master stability function framework to a broad class of degree-heterogeneous, weighted, and directed delay-coupled networks. The analysis reveals that synchronization can be enhanced in heterogeneous networks compared to homogeneous ones, including all-to-all networks, which are known to be optimal in non-delayed systems. To identify optimal heterogeneous structures, we develop a network optimization method that finds directional, edge-weighted configurations maximizing synchronization stability. Our results show that, in delay-coupled networks, heterogeneity and nonreciprocity are key resources for synchronization.  
\end{abstract}

%\keywords{Suggested keywords}%Use showkeys class option if keyword
                              %display desired
\maketitle

\textit{Introduction}\textemdash Finite signal propagation speeds and processing latencies make interactions in networked systems inherently time-delayed, introducing memory effects that fundamentally alter their dynamical behavior. Such delays arise in spike transmission across neural networks \cite{merolla2014million,stoelzel2017axonal,meszaros2025efficient} and light propagation in optically-coupled lasers \cite{zamora2010crowd,junges2013characterization,barioni2025interpretable,ye2025optimal}. They also manifest as communication lags in distributed learning \cite{li2020federated,liu2022decentralized,zhang2023understanding}, multi-agent control \cite{tian2008consensus,montanari2025optimal,trinh2026existence}, %,yu2010some
phase-locked circuits \cite{punetha2022heterogeneity}, and social group alignment \cite{shahal2020synchronization}.
These systems often rely on global synchronization to achieve coordination, coherence, or consensus. Yet, in the related area of oscillator networks, time delays have been shown to generally degrade synchronization, frequently leading to %complex dynamical behaviors such as chaos \cite{wernecke2019chaos}, 
oscillation death \cite{zou2011control,biswas2024transition}, %spiking \cite{Azangue2024Stability,Wang2024Synchronization}, 
attractor competition \cite{Kim1997Multistability,Williams2013Synchronization,kantner2015delay}, and chimera states \cite{sethia2008clustered,hart2016experimental,gjurchinovski2017control}.

\vspace{-6pt}
These challenges naturally raise the question of how to design network structures that promote synchronization stability in delayed systems. In the absence of delays, the master stability function (MSF) formalism provides a powerful framework for identifying such networks \cite{pecora1998master,acharyya2026comprehensive}. To reduce the dimensionality of the stability analysis, this formalism relies on three key assumptions: 1)~the intrinsic dynamics of all nodes are identical, 2)~\rev{the coupling term admits an invariant synchronization manifold}, and 3)~the coupling matrix is diagonalizable. These assumptions have since been relaxed to account for nondiagonalizable networks \cite{nishikawa2006synchronization,Nishikawa2006maximum}, cluster synchronization \cite{pecora2014cluster}, higher-order interactions \cite{mulas2020coupled,gambuzza2021stability,zhang2021unified,carletti2023global}, and non-identical oscillators \cite{sun2009master,zhang2018identical}. Yet, in the presence of time delays, the coupling function \rev{is generally non-diffusive} and hence the second MSF assumption often fails, rendering the design of synchronizable networks a fundamental challenge. Although prior extensions of the MSF analysis have considered delayed interactions \cite{choe2010controlling,keane2012synchronisation,dahms2012cluster,leng2016basin,borner2020delay,mahdavi2025time}, these extensions remain largely restricted to \rev{degree-homogeneous} network structures\textemdash which we show here to be \textit{suboptimal} for synchronization. Crucially, a general framework capable of characterizing heterogeneous network structures, including optimal ones, is still lacking for delayed systems. 

\begin{figure}%[tbhp]
\centering
\includegraphics[width=1\linewidth]{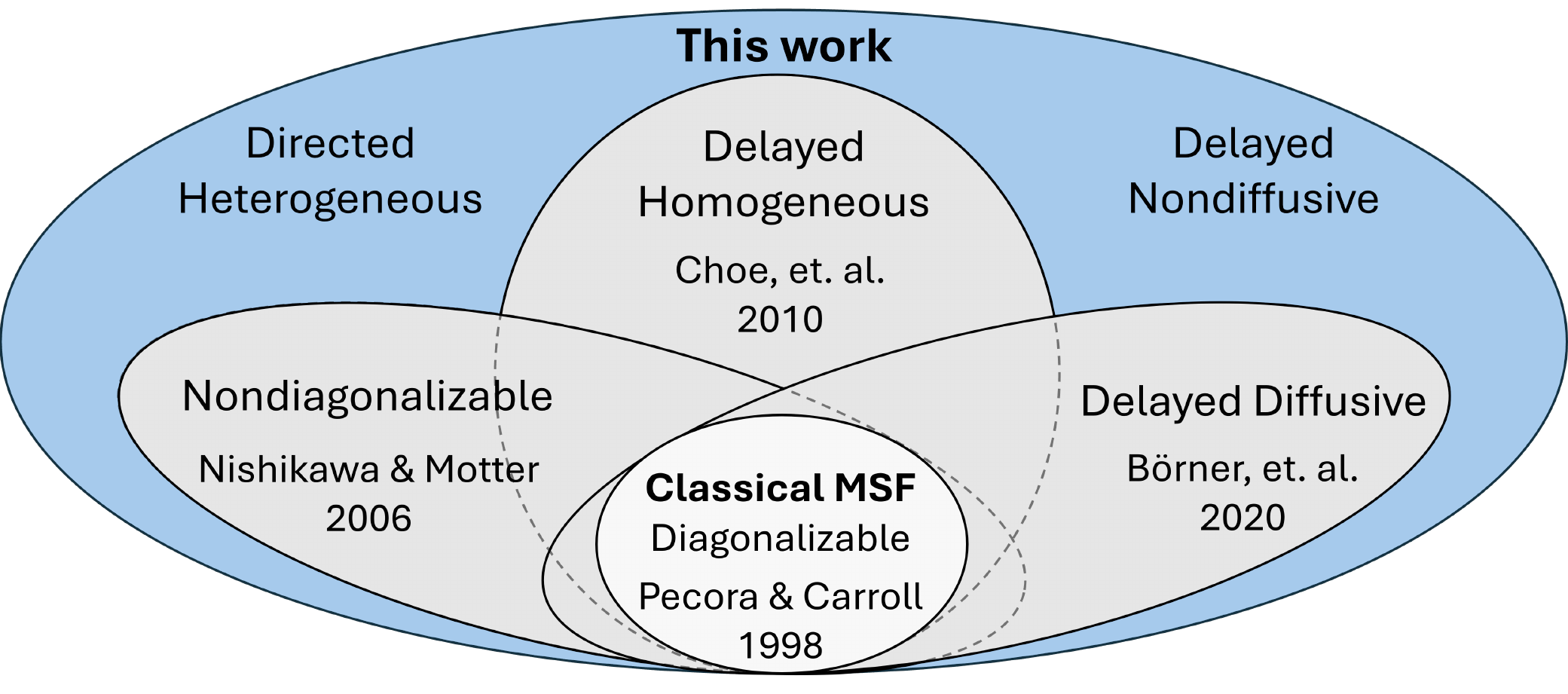}
\caption{Key extensions of the MSF framework to delay-coupled systems, where each set represents a methodological advance toward a broader class of network systems.}
\label{fig:diagram}
\vspace{-9pt}
\end{figure}

\pagebreak
In this Letter, we generalize the MSF framework to delay-coupled dynamics and propose an optimization method to identify networks that maximize synchronization stability. The extended framework applies to weighted, directed, and degree-heterogeneous networks, including diffusive and non-diffusive coupling schemes not accounted for by existing methods (Fig.~\ref{fig:diagram}). For both coupling types, time delays reduce the set of networks for which synchronization is stable by reshaping the stability region of the underlying MSF. We show that, in delayed systems, optimal synchronization stability requires the networks to be directed and have a heterogeneous distribution of indegrees. Moreover, in the small-delay regime, optimized networks can achieve substantially greater stability than in the non-delayed case. With phase-amplitude oscillators and coupled lasers as representative examples, these results establish heterogeneity as a principled mechanism for the emergence of stable collective behavior in delay-coupled networks.

%===================================================================================
\smallskip\noindent
\textit{MSF generalization for delay-coupled systems}\textemdash 
Consider a network of $M$ delay-coupled identical oscillators:
\begin{equation}
\dot{\textbf{x}}_j(t) = \textbf{f}\big(\textbf{x}_j(t)\big) + \sum_{k=1}^M A_{jk} \textbf{h}\big(\textbf{x}_j(t),\textbf{x}_k(t-\tau)\big),
\label{Dyn_LK_Eq}
\end{equation}
where $\textbf{x}_j \in \mathbb{R}^n$ denotes the state of oscillator $j$, function $\textbf{f}$ governs the intrinsic dynamics of each isolated oscillator, $\textbf{h}$ is the coupling function between the current state $\textbf{x}_j(t)$ and the delayed state $\textbf{x}_k(t-\tau)$, $A\in\mathbb R^{M\times M}$ is the (weighted) adjacency matrix encoding the network structure, and $\tau$ is the time delay.

Our main focus is the stability of the identical synchronous state $\mathbf{x}_1^*(t) = \ldots = \mathbf{x}_M^*(t) = \mathbf{x}^*(t)$. To express the corresponding variational equation, we define $\boldsymbol{\eta}_j(t)$ as the vector of small deviations from $\textbf{x}_j^*$ and collect them into the global perturbation vector $\boldsymbol{\eta}(t) = (\boldsymbol{\eta}_1^\top,\hdots,\boldsymbol{\eta}_M^\top)^\top$. Linearizing around $\textbf{x}^*$ yields
\begin{equation}
\begin{aligned}
\dot{\bm{\eta}}(t)=\left(\operatorname{D}^{(0)}\textbf{f}(\textbf{x}^*) \otimes I_M+ \operatorname{D}^{(0)}\textbf{h}(\textbf{x}^*) \otimes \Delta\right) \bm{\eta}(t)\\
    + \,  \operatorname{D}^{(\tau)}\textbf{h}(\textbf{x}^*) \otimes A \bm{\eta}(t-\tau),
\end{aligned}
\label{eq.msf.kronvector}
\end{equation}
where $\otimes$ denotes the Kronecker product, $I_{M}$ is the identity matrix of order $M$, and $\Delta=\operatorname{diag}\left(d_1,\ldots,d_M\right)$ is the diagonal matrix of node indegrees $d_j = \sum_k A_{jk}$. The operator $\operatorname{D}^{(\tau)}$ denotes the Jacobian of a vector field with respect to the delayed state $\textbf{x}(t-\tau)$; accordingly, $\operatorname{D}^{(0)}$ denotes the Jacobian with respect to $\textbf{x}(t)$.
Thus, the variational equation can be written in matrix form as
\begin{equation}
\begin{aligned}
    \dot{\boldsymbol{\eta}}(t) = {J}_1\rev{(t)} \, \boldsymbol{\eta}(t) + {J}_2\rev{(t)}
 \, \boldsymbol{\eta}(t-\tau),
\end{aligned}
   \label{VariationalEqs}
\end{equation}
\noindent
\rev{where ${J}_1(t)$ and ${J}_2(t)$ are evaluated along the synchronous trajectory $\mathbf x^*(t)$ and are therefore generally time dependent. When the synchronous solution is stationary or can be mapped to a co-rotating frame in which the Jacobians are time independent, Eq.~\eqref{VariationalEqs} reduces to a linear delay equation with constant matrices $J_1$ and $J_2$. In this case,} Eq.~\eqref{VariationalEqs} admits a solution of the form $\boldsymbol{\eta}(t) = \boldsymbol{\eta}(0)e^{\lambda_\ell t}$, whose stability exponents $\lambda_\ell$ are determined by the  characteristic equation \cite{bellen2013numerical,richard2003time}
\begin{equation}
\begin{aligned}
    \det\left({J}_1 + {J}_2 \, e^{-\lambda_\ell \tau} - \lambda_\ell {I}_{Mn}\right) = 0.
\end{aligned}
   \label{CharacteristicEq}
\end{equation}

\noindent
For $\tau\neq 0$, Eq.~\eqref{CharacteristicEq} admits infinitely many solutions $\lambda_\ell$. Notwithstanding, the synchronization stability is determined by the maximum transverse Lyapunov exponent (MTLE) $\lambda_{\rm max}(J_1,J_2) = \operatorname{max}_\ell\{\operatorname{Re}(\lambda_\ell)\}$.

To derive the MSF formalism for delay-coupled systems of the form \eqref{Dyn_LK_Eq}, we seek to separate the contribution of the network structure from that of the local oscillator dynamics, reducing the stability analysis from  $Mn$ to  $n$ dimensions.
%Consequently, the synchronous solution is itself determined by the delayed coupling and is not governed solely by the intrinsic dynamics $\mathbf{f}$.
%
We begin by defining the $n\times M$ matrix $\xi = [\bm\eta_1 \,\, \ldots \,\, \bm\eta_M]$, for which Eq.~\eqref{eq.msf.kronvector} can be equivalently written as
\begin{equation}
  \dot{\xi}(t)= \operatorname{D}^{(0)} \textbf{f} \,\xi(t)+  \operatorname{D}^{(0)}\textbf{h}\, \xi(t) \Delta+ \operatorname{D}^{(\tau)}\textbf{h} \,\xi(t-\tau) A^{\top} .
\label{eq.msf.matrixform}
\end{equation}
Note that the coupling in system~\eqref{Dyn_LK_Eq} is generally non-diffusive due to the presence of time delay. In particular, even when $\mathbf{h}\big(\mathbf x_j(t),\mathbf x_k(t-\tau)\big) = \mathbf x_k(t-\tau) - \mathbf x_j(t)$, the coupling function does not vanish for $\tau>0$, except when $\mathbf{x}^*$ is an equilibrium point or a periodic orbit such that $\tau$ is a multiple of the period. Consequently, the adjacency matrix $A$ and the degree matrix $\Delta$ in Eq.~\eqref{eq.msf.matrixform} cannot always be aggregated into a Laplacian matrix $L = \Delta-A$, precluding direct application of the standard MSF analysis \cite{pecora1998master}. We address this challenge by considering the change of variables $\zeta = \xi (P^{-1})^\top$, where $P\in\R^{M\times M}$ is a similarity matrix that \textit{simultaneously transforms}: i) $A$ into its Jordan canonical form $J = P^{-1} A P$ and ii) $\Delta$ into a lower triangular form $\widetilde \Delta = P^{-1} \Delta P$. 
Under this transformation, it follows from Eq.~\eqref{eq.msf.matrixform} that
\begin{equation}
\dot{\zeta}(t)= \operatorname{D}^{(0)} \textbf{f} \,\zeta(t)+  \operatorname{D}^{(0)}\textbf{h}\, \zeta(t) \widetilde \Delta^{\top}+ \operatorname{D}^{(\tau)}\textbf{h} \,\zeta(t-\tau) J^{\top},
\label{eq.msf.matrixform2}
\end{equation}
which defines an interdependent set of $M$ equations, one for each column $\bm\zeta_j$ of $\zeta$. Special classes of networks for which such a matrix $P$ exists\textemdash including all-to-all, ring, and master-slave networks\textemdash are presented in detail in the Supplemental Material (SM)~\cite{supplemental_mat}, Sec.~\ref{sm:ConditionsForTheory}.

\begin{figure*}[th]
\centering
\includegraphics[width=0.86\linewidth]{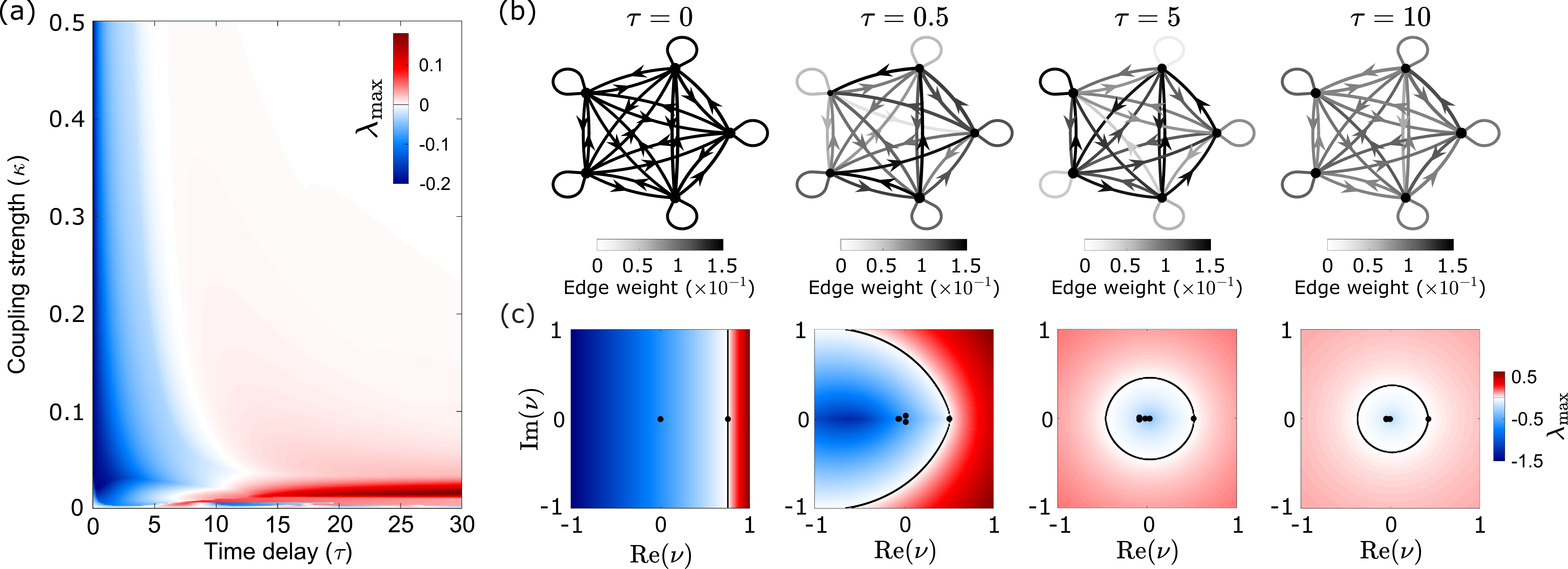}
\caption{MSF analysis of delay-coupled SL oscillators. 
(a)~MTLE of the synchronous state as a function of the coupling strength $\kappa$ and time delay $\tau$ for an all-to-all network configuration, where blue (red) regions indicate stable (unstable) synchronization. \rev{Discontinuities observed for small $\kappa$ arise from the multistability of the system, as further discussed in SM \cite{supplemental_mat}, Sec.~\ref{sm:Multistab}}.
(b,~c)~Optimal network configuration (b) and corresponding MSF landscape (c) for increasing $\tau$. In panel b, the edge thickness indicates the interaction weights $A_{ij}$. %and the node size is proportional to the weighted outdegree $\sum_i A_{ij}$, highlighting master nodes.
Panel c shows the stability region $\Omega$ in the complex plane (blue) and the eigenvalues of the corresponding adjacency matrix (dots), which lie entirely within or on the boundary of $\Omega$. 
%
%For $\tau = 10$, note that a homogeneous all-to-all network with $\kappa = 0.15$ has its eigenvalue $\nu=0.75$ outside $\Omega$ and is therefore unstable.
%
The SL model parameters are $( \omega,\lambda, \gamma) = (0.25,0.1,-4.4)$, and the MSFs are computed using the smallest indegree $d^\star$ of each optimal network. 
}
\vspace{-10pt}
\label{fig:SL_Opt}
\end{figure*}

To make contact with the previous literature, consider the particular case in which $A$ and $\Delta$ are \textit{simultaneously diagonalizable} by $P$, and thus the $M$ modes in Eq.~\eqref{eq.msf.matrixform2} decouple as in the original MSF:
\begin{equation}
    \dot{\bm\zeta_j}(t) = \left(\operatorname{D}^{(0)} \textbf{f} +  \widetilde{\Delta}_{jj} \operatorname{D}^{(0)}\textbf{h}\right) \bm\zeta(t) + \nu_j \operatorname{D}^{(\tau)}\textbf{h} \, \bm\zeta_j(t-\tau),
    \label{eq.msf.decomposed}
\end{equation}

\noindent
for $j=1,\ldots,M$. In this case, the stability of each orthogonal mode $\bm\zeta_j$ depends uniquely on the eigenvalue $\nu_j$ of the adjacency matrix $A$.
However, $A$ and $\Delta$ are generally not simultaneously diagonalizable. The only two scenarios that satisfy this property were considered previously in Ref.~\cite{choe2010controlling}: i) $A$ is itself diagonal (i.e., the network is uncoupled) and ii) $A$ is diagonalizable and $\Delta = d I_M$ (i.e., $d_j = d, \forall j$, and thus all indegrees are identical). 

Therefore, extending the MSF framework to a broader class of network structures requires the analysis of nondiagonalizable coupling matrices, as previously considered for non-delayed systems in Refs.~\cite{nishikawa2006synchronization,Nishikawa2006maximum}. Such matrices accommodate directed network structures, often described by heterogeneous degree distributions. Assuming that $J$ contains nontrivial Jordan blocks, each column of $\zeta$ is decoupled from those corresponding to other Jordan blocks. For a Jordan block of size $\ell\times \ell$ associated with the eigenvalue $\nu$, we have
\begin{equation}
\begin{aligned}
\dot{\bm\zeta}_1(t)&=\Xi_1, \\
\dot{\bm\zeta}_2(t)&=\Xi_2 + \operatorname{D}^{(0)}\textbf{h}  \,\,\widetilde{\Delta}_{21}  \bm\zeta_1(t)+ \operatorname{D}^{(\tau)}\textbf{h} \,\, \bm\zeta_1(t-\tau), \\
&\vdots \\
\bm\dot{\bm\zeta}_\ell(t)&=\Xi_\ell+ \operatorname{D}^{(0)}\textbf{h} \sum_{j=1}^{\ell-1} \widetilde{\Delta}_{\ell j} \bm\zeta_j(t)+ \operatorname{D}^{(\tau)}\textbf{h} \,\, \bm\zeta_{\ell-1}(t-\tau),
\end{aligned}
\label{eq.msf.jordanblock}
\end{equation}
\noindent
where the subsystem $\Xi_j$ of each mode $j\in\{1,\ldots,\ell\}$ is given by
\begin{equation}
\Xi_j = \underbrace{(\operatorname{D}^{(0)} \textbf{f}+\widetilde{\Delta}_{jj} \operatorname{D}^{(0)}\textbf{h} )}_{J_{1,j}'} \bm\zeta_j(t)+ \underbrace{\nu \operatorname{D}^{(\tau)}\textbf{h}}_{J_2'(\nu)} \,\bm\zeta_j(t-\tau).
\label{eq.subsystem}
\end{equation}
Equation~\eqref{eq.msf.jordanblock} has a staircase form due to the triangular structure of $J$ and $\widetilde \Delta$: the dynamics of each mode $\bm\zeta_j$ depend only on itself and the preceding modes $\bm\zeta_{j-k}$, for $k=1,\ldots,j-1$. Consequently, if subsystem $\dot{\bm\zeta}_j(t) = \Xi_j$ is stable, it follows that, in the full system \eqref{eq.msf.jordanblock}, $\bm\zeta_{j-1}(t)\rightarrow 0$ implies $\bm\zeta_j(t)\rightarrow 0$ as $t\rightarrow\infty$. Subsystem $\Xi_j$ is stable if $\lambda_{\rm max}(J_{1,j}',J_2')<0$, with the key observation that this MTLE can be computed from Eq.~\eqref{CharacteristicEq} using $n\times n$ rather than $Mn\times Mn$ matrices. By induction, the full system \eqref{eq.msf.jordanblock} is stable if $\lambda_{\rm max}(J_{1,j}',J_2')<0,\, \forall j$.

%%%%%%%%%%%%%%%%%%%%%%
Two analyses are pertinent: across Jordan blocks and across networks. Since the subsystems $\Xi_j$ present across Jordan blocks are identical, only the stability analysis of system \eqref{eq.msf.jordanblock} associated with the largest block is required. We thus focus on characterizing how the network structure, parametrized by $\nu$, influences the stability of system \eqref{eq.msf.jordanblock}. Let $\Omega_j = \{ \nu \in \C: \lambda_{\max}\big(J'_{1,j},J_2'(\nu)\big) < 0\}$ be the region of the complex plane for which the subsystem $\Xi_j$ is stable. From above, the stability region of system \eqref{eq.msf.jordanblock} is given by $\Omega = \bigcap_j \Omega_j$, which in principle requires the analysis of \textit{all} subsystems $\Xi_j$. In the special case of \textit{small delay} (i.e., $\tau\rightarrow 0$) and \textit{identity Jacobian coupling} (i.e., ${\rm D}^{(\tau)}\mathbf{h} = -{\rm D}^{(0)}\mathbf{h}= I_n$), we can prove that $\Omega_j$ expands monotonically with $\widetilde \Delta_{jj}$ (see SM \cite{supplemental_mat}, Sec.~\ref{sm:ConditionsForTheory}, where the MSF conditions are verified). 
Thus, if $\widetilde \Delta_{jj} < \widetilde \Delta_{kk}$, then $\Omega_j\subset\Omega_k$.  
Since the diagonal elements of $\widetilde D$ satisfy $\min_j d_j \leq \widetilde \Delta_{jj} \leq \max_j d_j$, it follows that, under these conditions, the stability region $\Omega$ is determined solely by the subsystem $\Xi_j$ associated with the smallest indegree $d^\star = \min_j d_j$ in the network. As shown below, the monotonic dependence of $\Omega_j$ on $\widetilde{\Delta}_{jj}$ is also observed numerically in the regime of large delays.
%
%For larger $\tau$, we note empirically that $\Omega_j$ still varies monotonically with $\widetilde D_{jj}$ depending on the coupling scheme, as discussed later. %either expanding or shrinking as $\widetilde D_{jj}$ increases (SM \cite{supplemental_mat}, Fig.~\ref{fig:varying_degrees}); in these cases, the stability region also reduces to a single set $\Omega = \Omega_{j}$, where $\Omega_j$ corresponds to the subsystem with either the smallest or the largest indegree (denoted by $d^\star$).

We now reach the following central conclusion: \textit{The synchronization stability of the delay-coupled system~\eqref{Dyn_LK_Eq} is entirely determined by the MSF function $\lambda_{\rm max}\big(J_1', J_2'(\nu)\big)$, which maps the complex eigenvalue $\nu$ associated with the adjacency matrix $A$ to the MTLE computed from the $n\times n$ matrices $J_1' = (\operatorname{D}^{(0)} \mathbf{f}+ d^\star \operatorname{D}^{(0)}\mathbf{h})$ and $J_2' = \nu \operatorname{D}^{(\tau)}\mathbf{h}$.} This result is relevant for network design, as it reduces the full stability problem to the spectral properties of $A$ and a low-dimensional MSF. 
Note, however, that the MSF itself is constructed for a prescribed minimal indegree $d^\star$. Therefore, when searching for adjacency matrices $A$ whose eigenvalues satisfy $\nu_j\in\Omega$, one must ensure that these eigenvalues correspond to networks that satisfy the structural constraint $d_j \geq d^\star$, $\forall j$.  Consequently, not every spectral configuration contained in $\Omega$ is necessarily admissible. 

%===================================================================================

\smallskip\noindent
\textit{MSF analysis for heterogeneous networks}\textemdash To illustrate our results, we consider a network of $M$ delay-coupled Stuart-Landau (SL) oscillators governed by
\begin{equation}
\dot{z}_j(t)=f\left(z_j(t)\right)+ \sum_{k=1}^{M} A_{jk} \big(z_k(t-\tau)-z_j(t)\big),
\label{eq:Dyn_SL}
\end{equation}
where the complex variable $z_j= r_j e^{i\phi_j}$ represents the state of oscillator $j$. The isolated oscillator dynamics follow the normal form close to a Hopf bifurcation:
\begin{equation}
f\left(z_j\right) = \big(\lambda +i \omega -(1+i\gamma)|z_j|^2\big)z_j,
\label{eq:LocalDynSL}
\end{equation}
where real parameters $\lambda$, $\omega$, and $\gamma$ characterize the amplitude growth rate, natural frequency, and amplitude-dependent frequency shift of the limit-cycle dynamics, respectively. Since oscillators are identical, the system admits identical synchronous solutions given by $z_j = r^*e^{i\Omega t}, \, \forall j$. \rev{We work in a co-rotating frame and focus on the stabilization of synchronous states associated with the smallest frequency shift $\Omega_{0}=\min |\Omega - \omega|$}. Substituting this solution into Eq.~\eqref{eq:Dyn_SL} in polar coordinates $(\dot r_j,\dot\phi_j)=(0,\Omega_0)$ yields a set of transcendental equations that can be solved numerically to identify $(r^*,\Omega_0)$. For the specific case of all-to-all networks (defined as $A_{jk} = \kappa$, $\forall \,j,k$), Fig.~\ref{fig:SL_Opt}(a) shows that the synchronous state becomes unstable as the delay $\tau$ increases, especially at intermediate values of coupling strength $\kappa$.

We now apply our MSF analysis to the SL model to systematically characterize the interplay between synchronization stability, network structure, and time delay [Fig.~\ref{fig:SL_Opt}(b), (c)]. For each choice of $\tau$, we consider an optimal network configuration that minimizes $\lambda_{\rm max}$ subject to the edge constraint $A_{jk} \le A_{\rm max}$. Details on the optimization procedure used to construct these networks and verification of the existence conditions of the similarity matrix $P$, are provided in SM \cite{supplemental_mat}, Sec.~\ref{sm:ConditionsForTheory}.
Synchronization stability can thus be directly verified through the MSF: in all optimal networks, the eigenvalue spectrum of $A$ lies entirely within the stability region $\Omega$, except for a single eigenvalue $\nu_0$ associated with the longitudinal mode, corresponding to $\lambda_{\rm max}(\nu_0)=0$. The example networks in Fig.~\ref{fig:SL_Opt}(b), (c) demonstrate that our formalism captures the stability properties of systems with heterogeneous degree distributions, directed edges, and self-interactions.

Figure~\ref{fig:SL_Opt}(c) further shows that increasing the time delay systematically shrinks the stability region of the MSF. The same network structures that are stable in the absence of delay may thus lose stability as $\tau$ increases. The effect is particularly evident for degree-homogeneous networks with $d_j=d, \, \forall j$, which have the eigenvalue $\nu_0 = d$ associated with the longitudinal mode. As $\Omega$ shrinks with increasing $\tau$, this eigenvalue eventually exits the stability region so that $\lambda_{\rm max}(\nu_0) > 0$, destabilizing the synchronous state. For example, in a 5-node all-to-all network with $\kappa = 0.15$, the eigenvalue $\nu_0= d = 0.75$ lies outside $\Omega$ for $\tau = 10$, rendering the corresponding mode unstable. (Because the coupling \rev{term does not generally vanish for $\tau > 0$ at the synchronous state $\mathbf x^*$}, the Lyapunov exponent $\lambda_{\rm max}(\nu_0)$ is not necessarily zero, and the longitudinal mode may become unstable.) Consequently, degree-homogeneous networks lose stability for sufficiently large $\tau$, making them suboptimal choices for synchronization in the presence of delays. 

%Because the SL oscillator is the universal normal form of a Hopf bifurcation, these conclusions hold for any system operating near the Hopf onset.

%===================================================================================
\medskip\noindent
\textit{MSF analysis for non-diffusive coupling}\textemdash
Although the \rev{coupling function of the SL model \eqref{eq:Dyn_SL} is generally non-diffusive when $\tau>0$, it reduces to a diffusive form in the limit $\tau=0$.} We now show that our MSF framework applies more generally to systems with intrinsically non-diffusive coupling for all $\tau\geq0$. As a representative example, we consider a network of $M$ coupled diode lasers described by the Lang-Kobayashi (LK) model \cite{lang1980external,masoller1997implications}:
\begin{equation}
\small{
\begin{aligned}
    \dot{E}_j(t)&=  \frac{1+i \alpha}{2}\left(G_j-\gamma \right) E_j(t)+ i\omega E_j(t)+ \sum_{k=1}^M {A}_{j k} E_k(t-\tau),
    \\
    \dot{N}_j(t) &=  J_{0}-\gamma_{n} N_j(t)-G_j\left|E_j(t)\right|^2,
\end{aligned}}
\label{eq.lk}
\end{equation}

\noindent
where $E_j(t)$ represents the electric field and $N_j(t)$ the carrier number of laser $j$, and $\tau$ is the round-trip propagation time of light. The frequency detuning $\omega$ is measured relative to the lasing-mode frequency $\omega_0$, set by the cavity length, and $\alpha$ is the linewidth enhancement factor accounting for the phase-amplitude coupling. The function $G_j(t)=g \big(N_j(t)-N_{0}\big)/\big(1+sr_j^2(t)\big)$ is the nonlinear optical gain, and $\{g,s,\gamma,\gamma_{n}, N_{0}, J_{0}\}$ are constructive parameters of the individual lasers.
To generate a coherent combined beam, all lasers must synchronize to the same frequency with minimal phase spreading to avoid destructive interference. Thus, we focus our stability analysis on stationary synchronous states of the form $E_j(t)=r^*e^{i\Omega_0 t}$ and $N_j(t)=N^*$, $\forall j$. Because $A_{jk}\geq 0$ for all entries $j,k$ (including self-coupling terms), the coupling term does not vanish for this stationary state even when $\tau = 0$. %Here, $\Omega$ is the frequency shift relative to the lasing mode frequency $\omega_0$\textemdash the ideal operating frequency\textemdash and our goal is to establish the stability conditions of such states.

\begin{figure}[tb]
\centering
\includegraphics[width=0.92\linewidth]{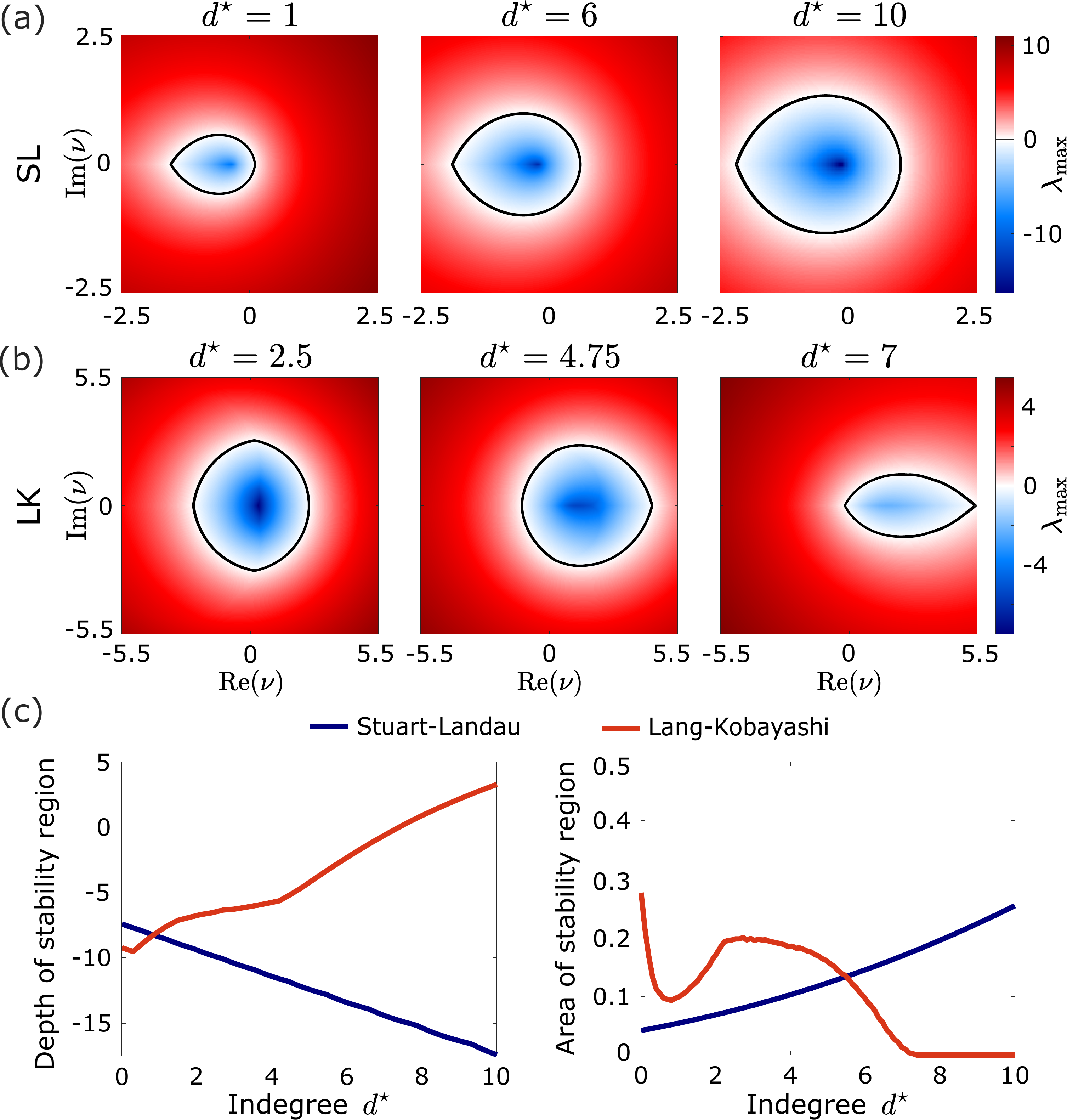}
\caption{MSF comparison analysis for the SL and LK models.
(a, b)~MSF landscape of the SL (a) and LK (b) models for $\tau = 0.1$, showing $\lambda_{\rm max}$ as a function of the tunable adjacency matrix eigenvalue $\nu$ for different indegrees $d^\star$. The stability region $\Omega$ is characterized by its area and depth (i.e., $\min_\nu \lambda_{\rm max}$). 
(c)~Dependence of the stability region's depth (left) and area (right) on $d^\star$ for the SL and LK models.
The SL model parameters are set as in Fig.~\ref{fig:SL_Opt}, while the LK model parameters are specified in SM \cite{supplemental_mat},~Sec.~\ref{sm:CouplingClasses}.
%
%The time delay is set to $\tau = 0.1$ for both models and the remaining system parameters are specified in the SM \cite{supplemental_mat},~Sec.~\ref{sm:CouplingClasses}.
}
\label{fig:DeepSize_compare}
\vspace{-15pt}
\end{figure}

%For non-delayed phase oscillator networks, stronger coupling (corresponding to higher indegree) enhances synchronization stability \cite{dorfler2013synchronization,nishikawa2017sensitive,Rodrigues2016kuramoto}. Time delays, however, can qualitatively alter this behavior. 

\begin{figure*}[ht]
\centering
\includegraphics[width=0.9\linewidth]{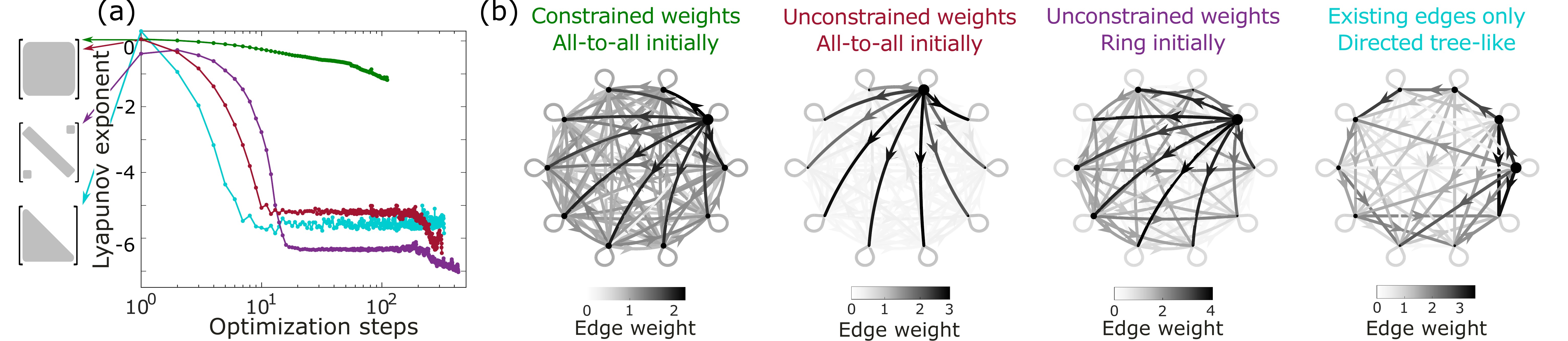}
\caption{Network optimization to enhance synchronizability.
(a)~Minimization of the MTLE of the LK model for different initial configurations and optimization constraints: all-to-all with fixed total weight (green); all-to-all with unconstrained weights (red); ring with unconstrained weights (purple); and directed tree-like structure with optimization restricted to existing edges and all others are set to zero (cyan). The structures of the initial adjacency matrices are shown on the left.
(b)~Resulting optimal network, where arrow direction and line thickness indicate edge directionality and weight, respectively; node size scales with outdegree, highlighting the most dominant nodes.
The model parameters are the same as in Fig.~\ref{fig:DeepSize_compare}.
}
\label{fig:Networks}
\vspace{-10pt}
\end{figure*}

Figure~\ref{fig:DeepSize_compare} contrasts the MSF landscapes of the SL and LK models. In the SL model, increasing $d^\star$ monotonically expands and deepens the stability region. This behavior is consistent with our theoretical predictions, since the delay is small and the SL model satisfies the identity Jacobian coupling. The LK model exhibits the opposite trend: increasing $d^\star$ destabilizes synchronization, as the stability region progressively shrinks and flattens, vanishing at $d^\star = 7.4$. Because in this model the state variable $N_j$ is not directly coupled to other oscillators (and hence ${\rm D}^{(\tau)}{\mathbf h}\neq I_n$), the monotonic dependence of $\Omega_j$ on $d_j$ is indeed not guaranteed to hold, and hence the MSF analysis of all subsystems $\Xi_j$ must be considered. In fact, both increasing $d^\star$ \textit{and} increasing $\tau$ tend to nonmonotonically reduce the size of the stability region $\Omega$ in the LK model. \rev{This loss of stability induced by large coupling is consistent with previous studies in semiconductor lasers \cite{heil1999influence,lenstra2003coherence}: due to strong phase-amplitude coupling (i.e., large $\alpha$), delayed interactions can generate amplitude fluctuations that feed back into the phase dynamics and amplify transverse perturbations, ultimately causing desynchronization.}
For more details and a comparison of the MSF for different classes of coupling schemes, see SM \cite{supplemental_mat}, Sec.~\ref{sm:CouplingClasses} and Fig.~\ref{fig:CouplingCompare}.

Crucially, for small $\tau$ and a range of indegrees $d^\star$, the stability region is smaller but deeper relative to the case $\tau = 0$ (SM \cite{supplemental_mat}, Fig.~\ref{fig:LEtau0}). This indicates that, for a given $d^\star$, appropriate choices of delay-coupled network configurations can achieve stronger synchronization stability than the optimal network choice for the non-delayed case.

\smallskip\noindent
\textit{Optimizing networks for synchronization}\textemdash 
%Since degree-homogeneous networks are suboptimal for synchronization in delay-coupled systems, 
The task of identifying optimal networks that maximize synchronization stability is generally nontrivial due to the constraints and dimensionality of the search space. To address this challenge, we develop an iterative optimization method that applies constrained weight perturbations $\delta A_{ij}$ to minimize the MTLE associated with a pre-specified synchronous solution $\textbf{x}^*$ (SM~\cite{supplemental_mat}, Sec.~SIV and Algorithm~\ref{alg.coupling_opt}). The optimization procedure terminates based on the specified maximum number of epochs and the optimization improvement tolerance.

Figure~\ref{fig:Networks}(a) shows the optimization performance under different constraints and initial conditions. 
% \textcolor{red}{Starting from an all-to-all topology and imposing the total-weight constraint $\sum_{i,j} \delta A_{jk} = 0$, the optimization yields a substantial stability improvement, with $\lambda_{\rm max}$ decreasing from $0.05$ to $-1.19$. 
% % while maintaining moderate weights ($0\leq A_{jk}\leq 2.3$). 
% Relaxing this constraint yields a further 491.4\% improvement relative to the constrained scenario, with only a 19.5\% increase in the maximum edge weight.}
% When initialized from a ring topology (which is inherently more stable than the all-to-all network), the unconstrained optimization achieves a final MTLE 9.0\% smaller than that obtained from the unconstrained all-to-all case. 
%
Across the three scenarios initialized by all-to-all and ring structures, the optimal configurations emerge as master-slave structures, characterized by predominantly directed interactions from nodes with large outgoing weights [Fig.~\ref{fig:Networks}(b)]. To take advantage of this structure, we introduce a fourth scenario in which the adjacency matrix is constrained to be lower triangular, explicitly enforcing a master-slave structure. This setting produces the most pronounced stability gains during the early optimization stages. 
Because the proposed algorithm outputs optimal configurations that are generically dense, we also investigate how \textit{sparsification} affects stability, revealing that some of the optimal configurations are strongly robust to edge trimming (SM \cite{supplemental_mat},~Sec.~\ref{sm:NetOpt}).

%===================================================================================
\smallskip\noindent
\textit{Conclusion}\textemdash Using phase-amplitude oscillator networks and coupled-laser arrays as representative systems, we show that time delays reshape synchronization stability across a broad range of coupling classes and network structures. Because we address phase-amplitude SL oscillators, the conclusions hold for any system operating near a Hopf bifurcation.
The numerical analysis reveals that optimal networks are characterized by master-slave (heterogeneous) configurations. \rev{This finding establishes a direct connection with studies of non-delayed network dynamics, which have shown that heterogeneous \cite{nishikawa2006synchronization,nishikawa2010network}, sparse \cite{mikaberidze2025emergent,ye2025optimal}, and directed \cite{son2009dynamics,zeng2011enhancing} networks can enhance synchronization under structural constraints. It is also} particularly relevant for engineering applications, where master-slave topologies are often preferred or presumed in decentralized control and communication. More broadly, our master stability framework can facilitate the stability analysis and design of synchronizable networks in systems with intrinsic delays, including for multi-agent control, federated learning, photonic communication, and neuromorphic computing.

%=======================================================
\textit{Acknowledgements}\textemdash The authors acknowledge support from the U.S.\ National Science Foundation (Grant No.\ DMS-2308341) and Office of Naval Research (Grant No.\ N00014-22-1-2200).

\textit{Data availability}\textemdash The code and data that support the findings of this article are openly available \cite{Data_avail}.

\let\oldaddcontentsline\addcontentsline% Store \addcontentsline
\renewcommand{\addcontentsline}[3]{}% Make \addcontentsline a no-op

% \bibliography{ref}

%apsrev4-2.bst 2019-01-14 (MD) hand-edited version of apsrev4-1.bst
%Control: key (0)
%Control: author (8) initials jnrlst
%Control: editor formatted (1) identically to author
%Control: production of article title (0) allowed
%Control: page (0) single
%Control: year (1) truncated
%Control: production of eprint (0) enabled
%

% MSF for delayed neural spiking
% https://www.science.org/doi/full/10.1126/science.1254642
% https://www.nature.com/articles/srep23471

% MSF for distributed optimization
% https://epubs.siam.org/doi/full/10.1137/22M1475648

\let\addcontentsline\oldaddcontentsline
%-------------------------------------------------------------------------------------------------
\clearpage
\onecolumngrid
\renewcommand{\thepage}{\arabic{page}}
\setcounter{page}{1}
\renewcommand{\thefigure}{S\arabic{figure}}
\renewcommand{\thetable}{S\arabic{table}}
\setcounter{figure}{0}
\setcounter{secnumdepth}{3}
\renewcommand{\thesection}{S\Roman{section}}
\setcounter{section}{0}
\renewcommand{\theequation}{S\arabic{equation}}
\setcounter{equation}{0}
\begin{center}
    \textbf{\large SUPPLEMENTAL MATERIAL \\[0.2cm]
    Generalized Master Stability of Heterogeneous Delay-Coupled Networks}
\end{center}

\vspace{-0.5cm}

\tableofcontents

\vspace{-0.3cm}

%=======================================================

\section{Conditions for the MSF of delay-coupled systems}
\label{sm:ConditionsForTheory}
In this section, we analyze the conditions underlying the theory.

%==================================
\bigskip\noindent
\textbf{Conditions on the network structure.} As discussed in the main text, the derivation of Eq.~\eqref{eq.msf.jordanblock} requires the existence of a similarity matrix $P$ that simultaneously transforms $A$ into its Jordan form and $\Delta$ into a lower-triangular matrix. We now prove that there are three classes of networks for which this simultaneous Jordanization and triangularization is always possible:

%==================================
\medskip\noindent
\textit{1) Networks with identical indegrees}\textemdash In this case, $\Delta = dI_n$. Therefore, for any similarity matrix $P$ (including one that transforms $A$ into its Jordan form), we have that $P^{-1} \Delta P = dP^{-1} I_n P = \Delta$, which is diagonal and hence lower triangular. Special cases of this class of networks include all-to-all and ring networks.

%==================================
\medskip\noindent
\textit{2) Master-slave networks with strictly heterogeneous self-coupling}\textemdash 
A master-slave network is defined by a directed graph with a single master node and a hierarchical organization among the remaining nodes. These networks are generally represented by the following lower-triangular adjacency matrix:
\begin{equation}
\begin{aligned}
A &= \begin{bmatrix}
A_{11} & 0  & \hdots & 0\\
A_{21} & A_{22}  & \hdots & 0\\
% A_{31} & A_{32} & A_{33} & \hdots& 0\\
\vdots & \vdots  & \ddots & 0\\
A_{M1} & A_{M2} & \hdots & A_{MM}
\end{bmatrix}.
\end{aligned}
\label{eq.sm.masterslaveadj}
\end{equation}
Here, we consider the case in which all the diagonal elements are strictly different, that is, $A_{ii}\neq A_{jj}$ for every $i\neq j$. Since $A$ is triangular, its eigenvalues are $\lambda_i = A_{ii}$, for $i=1,\ldots,M$, and they are all simple by assumption. Let $\mathbf p^{(i)}$ be an eigenvector associated with $\lambda_i$, and form the similarity matrix $P=[\mathbf p^{(1)} \,\, \cdots \,\, \mathbf p^{(M)}]$.

We first prove that the similarity matrix $P$ is lower triangular. Given that the eigenvector $\mathbf{p}^{(i)}$ satisfies $(A-\lambda_i I_M) \mathbf{p}^{(i)} = 0$ and that $A-\lambda_i I_M$ is lower triangular, we examine the eigenvalue equation row by row:

\begin{itemize}[noitemsep, topsep=0pt]
\item For $j<i$, we have $(A_{jj} - \lambda_i){p}_j^{(i)} = 0$ because all entries above the diagonal are zero. Since the eigenvalues are distinct, $A_{jj}-\lambda_i\neq 0$ and hence ${p}_j^{(i)} = 0$.

\item For $j = i$, we have $(A_{ii} - \lambda_i)\mathbf{p}_i^{(i)} + \sum_{k<i} A_{ik} {p}_k^{(i)} = 0$. Since $\lambda_i = A_{ii}$ and ${p}_k^{(i)} = 0$, $\forall k<i$, the equation is trivially satisfied for any choice of ${p}_i^{(i)}$. We thus normalize it as ${p}_i^{(i)}=1$.

\item For $j>i$, we have $\sum_{k\leq j} A_{jk}{p}_k^{(i)} - \lambda_i {p}_j^{(i)} = 0$. Since ${p}_k^{(i)} = 0$, $\forall k<i$, it follows that $(A_{jj} - \lambda_i) {p}_j^{(i)} = - \sum_{k=1}^{j-1} A_{jk} {p}_k^{(i)}$. Given that the eigenvalues are distinct, $A_{jj}-\lambda_i \neq 0$, and so we can recursively determine
\begin{equation}
    {p}_j^{(i)} = - \frac{ \sum_{k=1}^{j-1} A_{jk} {p}_k^{(i)} }{ A_{jj} - \lambda_i }.
\end{equation}
\end{itemize}

\noindent
Thus, the similarity matrix $P$ is lower triangular and hence so is its inverse $P^{-1}$. Because $\Delta$ is diagonal, $\Delta P$ is lower triangular, as each row of $P$ is scaled by $d_{ii}$. The product of lower triangular matrices is lower triangular, and thus $\widetilde{\Delta} = P^{-1} \Delta P$ is lower triangular, completing the proof.

%==================================
\medskip\noindent
\textit{3) Master-slave networks with identical self-coupling}\textemdash Consider the adjacency matrix \eqref{eq.sm.masterslaveadj} with identical nonzero diagonal elements, which we normalize to one without loss of generality: $A_{ii}=1$, $\forall i$. In this case, $A$ has repeated eigenvalues. Thus, it may not be diagonalizable, giving rise to nontrivial Jordan blocks. We denote the similarity matrix $P = [\mathbf{p}^{(1)} \,\, \ldots \,\, \mathbf{p}^{(M)}]$, whose columns satisfy a backward recursion of the form
\begin{equation}
\begin{aligned}
\mathbf{p}^{(i)} = \sum_{j=1}^{i-1} \mathbf{p}^{(i)}_{M-j}\,\mathbf{e}_{M-j}, \,\, \text{where} \,\,
\mathbf{p}^{(i)}_{M-j} = \frac{1}{A_{M-j+1,M-j}} \left(\mathbf{p}^{(i-1)}_{M-j+1} - \sum_{k=j+1}^{i-1} A_{M-j+1,M-k}\mathbf{p}^{(i)}_{M-k}\right)
\end{aligned}
\end{equation}
and $\mathbf{e}_i$ is the canonical basis vector. 
Given that $A_{i,i-1}\neq 0$, $\forall i$, it follows that the only nonzero entries $P_{ij}$ occur for indices $i>j$. Thus, $P$ is lower triangular and, following the steps described before, $\widetilde{\Delta}$ is also lower triangular.

%==================================
\medskip\noindent
\textit{Analysis for the optimal network structures}\textemdash Although we have rigorously proven the existence of a similarity matrix $P$ for the specific classes of adjacency matrices considered above, we now relax those structural conditions and provide numerical evidence that there can still exist a matrix $P$ that simultaneously transforms $A$ into its Jordan form and \textit{approximately} transforms $D$ into a lower triangular matrix, even when $A$ is neither lower triangular nor has homogeneous degrees. To illustrate this, we consider the optimal SL networks shown in Fig.~\ref{fig:SL_Opt}(b). These networks were obtained using our network optimization procedure (Algorithm~\ref{alg.coupling_opt}), initialized from the all-to-all network and subject to the constraints $0\leq A_{ij}\leq A_{\rm max}$, with $A_{\rm max} = 0.15$.

To assess whether $P$ satisfies the required conditions, we first find the transformation matrix $P$ that transforms the optimal adjacency matrix into its Jordan form, and then we calculate $\widetilde \Delta = P^{-1} \Delta P$. For each optimal network, we extract the lower triangular part of this matrix, $\widetilde \Delta^{\rm lower} = \left\{ \widetilde \Delta_{ij} \;:\; 1 \le i < j \le n \right\}$, and compute its Frobenius norm as \rev{$\left\lVert \widetilde \Delta^{\rm lower} \right\rVert_F = \frac{1}{M^2 \,\left\lVert A\right\rVert_2}\sqrt{\sum_{i,j}{\left(\widetilde \Delta^{\rm lower}_{ij}\right)^2}}$}. \rev{For comparison across network structures, note that we normalize the Frobenius norm by the network size $M$ and the spectral norm of the adjacency matrix.} 

For the optimal SL networks considered in Fig.~\ref{fig:SL_Opt}(b), (c), we obtained the following norms:
\begin{equation}
\begin{aligned}
\left\lVert \widetilde \Delta^{\rm lower} \right\rVert_F = \rev{0, \,\, 5.2\times 10^{-3}, \,\, 2.4\times 10^{-2}, \,\, \text{and} \,\, 2.7\times 10^{-2}}
\end{aligned}
\end{equation}
for time delays $\tau = 0$, $0.5$, $5$, an $10$, respectively. These results show that the deviation of the computed matrix $\widetilde \Delta$ from a lower triangular form remains small across the considered time delays. This supports the conclusion that our theoretical framework provides a reliable approximation of the MSF landscape for these network structures.

For the optimal LK networks in Fig.~\ref{fig:Networks}(b), we have
\begin{equation}
\begin{aligned}
\left\lVert \widetilde \Delta^{\rm lower} \right\rVert_F = \rev{2.5\times 10^{-2}, \,\, 3.0\times 10^{-3}, \,\, 6.5\times 10^{-3}, \,\, \text{and} \,\, 0}
\end{aligned}
\end{equation}
for the constrained all-to-all network, unconstrained all-to-all network, unconstrained ring network, and lower-triangular network, respectively.

%==================================
\bigskip\noindent
\textbf{Conditions on the monotonic variation of the stability region.} 
The stability of the eigenmodes in Eq.~\eqref{eq.msf.jordanblock} depends on the stability of all subsystems $\Xi_j$, for $j=1,\ldots,M$. Here, we show that, under certain conditions, the size of the stability region $\Omega_j\subset \C$ expands monotonically with $\widetilde \Delta_{jj}$. Consequently, the MSF analysis reduces to the stability problem of a single subsystem $\Xi_j$ associated with the smallest indegree $d^\star$.

To prove this, we consider the limiting case of small delay (i.e., $\tau \rightarrow 0$) and reciprocal coupling among all internal variables (i.e., ${\rm D}^{(\tau)}\mathbf{h} = - {\rm D}^{(0)}\mathbf{h} = I_n$). The stability of the time-delay system is determined by the generalized characteristic equation
\begin{equation}
    \det (J'_{1,j} + J_2' e^{\lambda_{\ell}\tau} - \lambda_{\ell}I_n) = 0,
\end{equation}
where $J'_{1,j} = {\rm D}^{(0)}\mathbf{f} - \widetilde \Delta_{jj} I_n$ and $J_2 = \nu I_n$. For small $\tau$, we can use the Taylor expansion $e^{\lambda_{\ell}\tau}\approx 1 - \lambda \tau$. Substituting this approximation yields the characteristic polynomial equation of a square matrix:
\begin{equation}
    \det \left({\rm D}^{(0)}\mathbf{f} + (\nu - \widetilde{\Delta}_{jj})I_n - \mu I_n \right) = 0,
\end{equation}
where $\mu = \lambda_{\ell}(1+\nu\tau)$. As $\widetilde{\Delta}_{jj}$ increases, we have that $\mu$ decreases, which implies that the entire spectrum $\lambda_{\ell}$ is uniformly shifted leftward in the complex plane, making the system $\Xi_j$ more stable. As a result, if $\widetilde {D}_{jj} < \widetilde D_{kk}$, then $\Omega_j\subset \Omega_k$.

For the SL model with $\tau = 0.1$, Figs.~\ref{fig:DeepSize_compare}(a) and \ref{fig:CouplingCompare}(e) numerically confirm that the stability regions expand monotonically with the increase of the indegree $d^\star$. This result is expected given that the coupling matrices $D^{(0)}\mathbf{h}$ and $D^{(\tau)}\mathbf{h}$, reported in Eq.~\eqref{eq.sm.SLDfDh}, satisfy the conditions above: for sufficiently small $\tau$, $\sin(-\Omega_0\tau)\approx 0$ and $\cos(-\Omega_0\tau)\approx 1$, yielding  ${\rm D}^{(\tau)}\mathbf{h} \approx - {\rm D}^{(0)}\mathbf{h} \approx I_n$. On the other hand, Figs.~\ref{fig:DeepSize_compare}(b) and \ref{fig:CouplingCompare}(g) show that these results do not hold for the LK model, which is also in agreement with our theory given that ${\rm D}^{(\tau)}\mathbf{h}$ and ${\rm D}^{(0)}\mathbf{h}$, reported in Eq.~\eqref{eq.sm.LKDfDh} are not proportional to the identity matrix due to the lack of inter-coupling between the carrier number variables.

%==================================
\bigskip \noindent
\textbf{Existence conditions of an identical stationary solution.} A key assumption for the MSF analysis is the existence of an identical stationary solution $\mathbf x^*$ across all nodes.
%First, our theory assumes that although the indegrees $d_j$ may vary across nodes, corresponding solutions of Eq.~\ref{eq.msf.kronvector} remain approximately identical across nodes so that the standard MSF dimension reduction remains valid. 
However, \rev{since the coupling terms in the SL and LK model do not generally vanish at the stationary state $\mathbf x^*$ for $\tau > 0$,} such an identical stationary solution may not exist depending on the choice of model parameters, time delay, and heterogeneity in the network structure. We thus also relax this assumption of exact identity and instead consider the existence of a near-identical stationary solution: $\mathbf{x}_j^*\approx \mathbf{x}_k^*$, $\forall (j,k)$, where $\mathbf{x}_j^*$ denotes the stationary solution of oscillator $j$. To assess the assumption, we compute the coefficient of variation $\epsilon_{\rm CV} = \frac{\rm{std} \{\mathbf{x}^*_j\}}{\rm{mean}\{\mathbf{x}^*_j\}}$ for all optimized networks shown in Fig.~\ref{fig:SL_Opt} and Fig.~\ref{fig:Networks}.

For the SL model (Fig.~\ref{fig:SL_Opt}), we obtain
$\epsilon_{\rm CV} = 0, \,\, 7.5\times 10^{-4}, \,\, 4\times 10^{-3}, \,\, \text{and} \,\, 3\times 10^{-2}$ for time delays $\tau = 0$, $0.5$, $5$, and $10$, respectively.
For the LK model (Fig.~\ref{fig:Networks}), we obtain $\epsilon_{\rm CV} = 5.6\times 10^{-4}, \,\, 1.8\times 10^{-4}, \,\, 1.8\times 10^{-4}, \,\, \text{and} \,\, 2.1\times 10^{-4}$ for the constrained all-to-all network, unconstrained all-to-all network, unconstrained ring network, and lower-triangular network, respectively.
The very small coefficients of variation for the SL and LK models support our assumption. Moreover, the numerical results presented in Fig.~\ref{fig:SL_Opt}(c) demonstrate good agreement with the theoretical predictions, as the eigenvalues of the optimized networks lie within the MSF stability region.

\begin{figure}[b]
\centering
\includegraphics[width=1\linewidth]{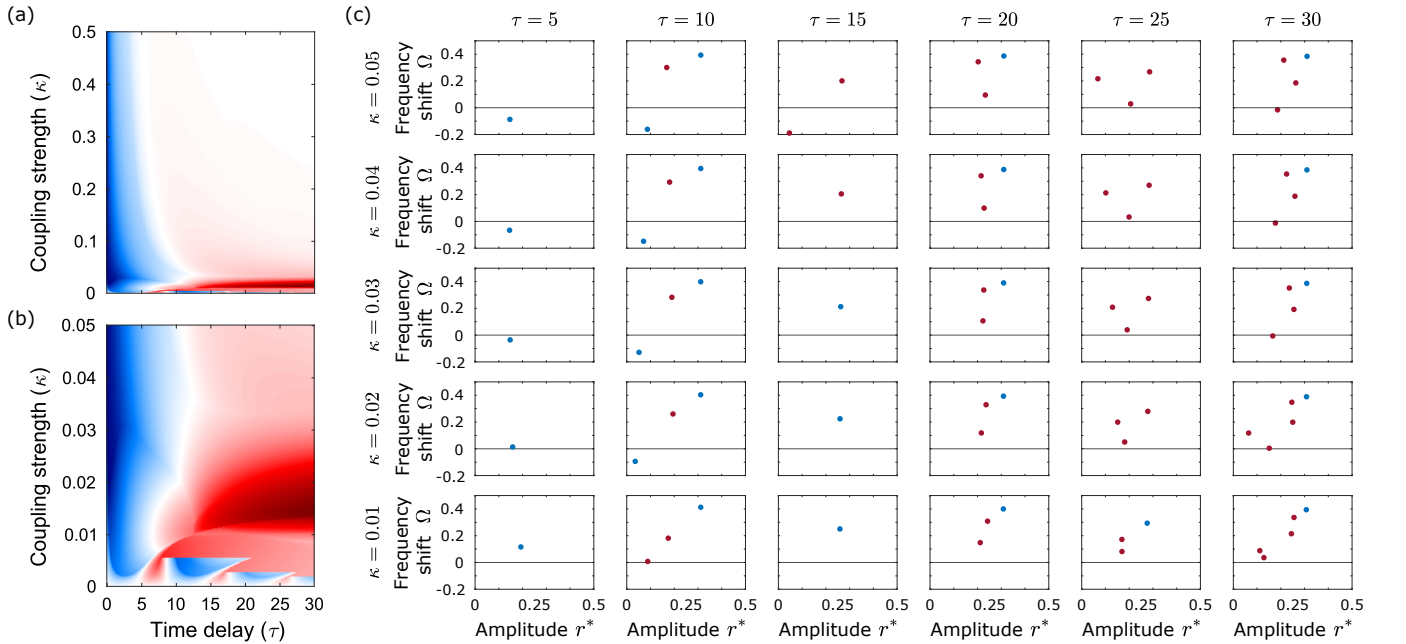}
\caption{\rev{Multistability of SL oscillator networks for varying levels of coupling strength and time delay. 
(a)~MTLE of the synchronous state as a function of the coupling strength $\kappa$ and time delay $\tau$ for an all-to-all network with $M=5$ nodes, as plotted in Fig.~\ref{fig:SL_Opt}. The blue and red regions indicate stable and unstable synchronization regions.
(b)~ Zoomed-in version of panel a for small $\kappa$.
(c)~Stationary synchronous states ($z_j(t) =r^*e^{i\Omega t},\, \forall j$) for increasing $\tau$ (left to right panels) and $\kappa$ (top to bottom panels). The stable and unstable states are marked in blue and red, respectively.
In panels (a) and (b), the stability is computed for the synchronous state corresponding to the smallest frequency shift $\Omega_{0}$.}
\label{fig:multistability_delay}}
\end{figure}

%=======================================================
\vspace{-0.4cm}
\rev{\section{Multistability analysis of the coupled Stuart-Landau model}
\label{sm:Multistab}}

\rev{The stationary synchronous states of the delayed SL model have the form $r_j(t) = r^*$ and $\phi_j(t) = \Omega t$ for all oscillators $j$ (i.e., same amplitude and frequency). For a fixed pair $(\tau, \kappa)$, this system generally admits multiple solutions, each defining a synchronous branch with a distinct amplitude $r^*$ and frequency $\Omega$.} 
\rev{Figure~\ref{fig:multistability_delay} illustrates this multistability by showing the coexisting solutions in the $(r^*,\Omega)$ plane. In the parameter range considered, both the number and location of solutions depend much more strongly on $\tau$ than on $\kappa$.}
\rev{Throughout the main text, including Fig.~\ref{fig:SL_Opt}(a) and Fig.~\ref{fig:multistability_delay}(a), (b), we specifically track the stability of the branch corresponding to the smallest frequency shift $\Omega_0 = \min|\Omega-\omega|$. This choice identifies the branch that connects continuously to the uncoupled limit. As $\tau$ increases, distinct branches can cross in frequency, potentially causing a different coexisting solution to become $\Omega_0$. Such branch switching may cause a discontinuous change in the MTLE associated with $\Omega_0$, as observed in Fig.~\ref{fig:multistability_delay}(b) for small $\kappa$.}

%=======================================================
 \bigskip\bigskip
\section{Classes of coupling schemes}
\label{sm:CouplingClasses}

We compare the results of our MSF analysis for different classes of diffusive and non-diffusive coupling schemes (summarized in Table~\ref{tab:coupling_classes}). Throughout, we set the SL model parameters as $( \omega,\lambda, \gamma) = (0.25,0.1,-4.4)$.

\begin{table}[H]
\footnotesize
\centering
\setlength{\tabcolsep}{6pt}
\renewcommand{\arraystretch}{1.25}

\begin{tabular}{p{1.9cm} p{3.2cm} p{7.8cm} p{3.0cm}}
\toprule[1.5pt]
\textbf{Type} &
\textbf{Class} &
\textbf{Coupling function} &
\textbf{Example} \\
\midrule[1.5pt]

%------------- Diffusive ------------------
\multirow{2}{*}{\centering\textbf{Diffusive}}
&
Non-delayed diffusive
&
$\mathbf{h}\!\left(\mathbf{x}_j(t),\mathbf{x}_k(t)\right)\,\,{\rm s.t.}\,\,
\mathbf{h}\!\left(\mathbf{x}(t),\mathbf{x}(t)\right)=0$
&
$\mathbf{x}_k(t)-\mathbf{x}_j(t)$
\\[3pt]

&
Delayed diffusive
&
$\mathbf{h}\!\left(\mathbf{x}_j(t-\tau),\mathbf{x}_k(t-\tau)\right) \,\,{\rm s.t.}\,\,
\mathbf{h}\!\left(\mathbf{x}(t-\tau),\mathbf{x}(t-\tau)\right)=0$
&
$\mathbf{x}_k(t-\tau)-\mathbf{x}_j(t-\tau)$
\\

\addlinespace[3pt]
\midrule[1.5pt]
\addlinespace[3pt]

%------------- Non-diffusive ------------------
\multirow{2}{*}{\centering\textbf{Non-diffusive}}
&
Delayed weakly diffusive
&
$\mathbf{h}\!\left(\mathbf{x}_j(t),\mathbf{x}_k(t-\tau)\right) \,\,{\rm s.t.}\,\,
\mathbf{h}\!\left(\mathbf{x}^*(t),\mathbf{x}^*(t-\tau)\right)=0$
for some $\mathbf{x}^*$
&
$\mathbf{x}_k(t-\tau)-\mathbf{x}_j(t)$
\\[3pt]

&
General delayed
&
$\mathbf{h}$ does not satisfy the conditions above for $\tau>0$
&
$\mathbf{x}_k(t-\tau)$
\\

\bottomrule[1.5pt]
\end{tabular}

\caption{Classification of coupling functions. We omit the condition
$\mathbf{h}\!\left(\mathbf{x}_k,\mathbf{x}_j\right)
= -\mathbf{h}\!\left(\mathbf{x}_j,\mathbf{x}_k\right)$ often included
in the definition of diffusive coupling because it is not necessary
to distinguish the classes above.}
\label{tab:coupling_classes}

\end{table}

\medskip\noindent
\textbf{Non-delayed diffusive coupling.} We start with the network system \eqref{Dyn_LK_Eq} in the non-delayed case $\tau = 0$, as originally considered in the classical MSF analysis by Pecora and Carroll \cite{pecora1998master}. Assume a non-delayed diffusive coupling, such as $\textbf{h}'(\textbf{x}_k(t)-\textbf{x}_j(t))$. In this case, it is well known that the coupling function vanishes at the identical state $\mathbf{x^*}$ and hence the variational equation \eqref{eq.msf.kronvector} can be decomposed as 
\begin{equation}
\begin{split}
    \dot{\bm\zeta_j}(t) = \operatorname{D}^{(0)} \textbf{f} \,\, \bm\zeta_j(t) + \nu \operatorname{D}^{(0)}\textbf{h} \,\, \bm\zeta_j(t).
\end{split}
\label{eq.nondelayedVarEqDiff}
\end{equation}

\begin{figure*}%[tbhp]
\centering
\includegraphics[width=0.85\linewidth]{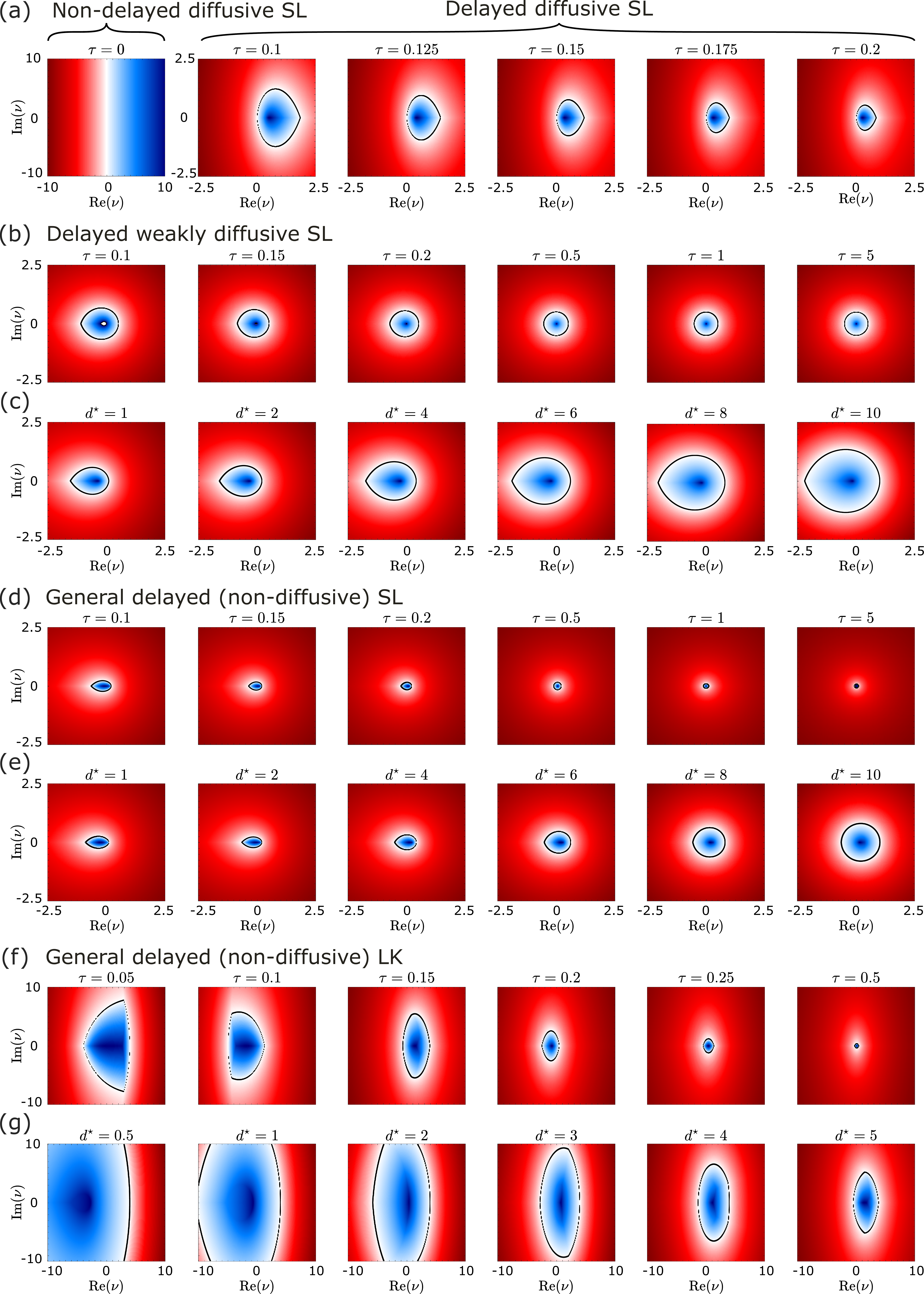}
\caption{Comparison between MSF landscapes for different coupling classes. 
(a--e) MSF analysis of the SL model with non-delayed and delayed diffusive coupling for increasing values of $\tau$ (a),
delayed weakly diffusive coupling for increasing values of $\tau$ (b) and $d^\star$ (c),
and \rev{general delayed (non-diffusive)} coupling for increasing values of $\tau$ (d) and $d^\star$ (e).
(f, g) MSF analysis of the LK model with \rev{general delayed (non-diffusive)} coupling for increasing values of $\tau$ (f) and $d^\star$ (g).
The color range of the MTLE values is normalized in each panel. Note that $\nu$ represents the eigenvalue of the Laplacian matrix for the diffusive cases in panel (a), while for all others it represents the eigenvalues of the adjacency matrix.}
\label{fig:CouplingCompare}
\end{figure*}

Taking the non-delayed coupled SL model as an example, we consider the dynamical equations in polar form 
\begin{equation}
\begin{aligned}
\dot{r}_j&=\left[\lambda-r_j^2\right] r_j+\sum_{k=1}^M A_{j k}\left[r_k \cos \left(\phi_k-\phi_j\right)-r_j\right], \\ 
\dot{\phi}_j&=\omega-\gamma r_j^2+ \sum_{k=1}^M A_{j k}\left[\frac{r_k}{r_j} \sin \left(\phi_k-\phi_j\right)\right].
\end{aligned}
\end{equation}
The identical synchronous solution $\mathbf{x}^*$ is given by $z(t) = \sqrt{\lambda}\,e^{i\left(\omega-\gamma\lambda\right)t}$. Linearizing around this solution, we obtain the matrices
\begin{equation}
\begin{aligned}
\operatorname{D}^{(0)}\textbf{f}  = \begin{bmatrix}
-2\lambda & 0\\
-2\gamma \sqrt{\lambda} & 0\\
\end{bmatrix}
\,\,\, \text{and} \,\,\,
\operatorname{D}^{(0)}\textbf{h} = 
I_2.
\end{aligned}
\end{equation}
The synchronous state stability is thus determined by the characteristic equation 
\begin{equation}\det\left(\operatorname{D}^{(0)}\textbf{f} - \left(\lambda_\ell - \nu \right) I_2\right) = 0.
\end{equation}
The corresponding MSF landscape, characterized by the Lyapunov exponent $\lambda_{\rm max} = \max_{\ell} {\rm Re}\{\lambda_{\ell}\}$, is illustrated in Fig.~\ref{fig:CouplingCompare}(a). It follows that the eigenvalues $\lambda_{\ell}$ of the system \eqref{eq.nondelayedVarEqDiff} are determined by $\lambda_\ell = \lambda_d + \nu$, separating the contribution of the oscillator dynamics (represented by the eigenvalue $\lambda_d$ of ${\rm D}^{(0)}_{\mathbf x}\mathbf{f}$) and the network structure (represented by the eigenvalue $\nu$ of the Laplacian matrix $L$). Clearly, the spectrum of the SL is uniformly shifted by the Laplacian eigenvalue $\nu$.

%==================================
\bigskip\noindent
\textbf{Delayed diffusive coupling.} Consider a system of $M$ oscillators with delayed coupling of the form
\begin{equation}
\dot{\textbf{x}}_j(t) = \textbf{f}(\textbf{x}_j(t)) + \sum_{k=1}^M A_{jk} \textbf{h}(\textbf{x}_j(t-\tau),\textbf{x}_k(t-\tau)).
\label{DynEqDelayDiff}
\end{equation}
This class of systems allows for diffusive coupling of the form $\textbf{h}\left(\textbf{x}_j(t-\tau),\textbf{x}_k(t-\tau)\right) = \textbf{h}'(\textbf{x}_k(t-\tau)-\textbf{x}_j(t-\tau))$, which vanishes at the identical state $\mathbf{x}^*$. This allows us to write the variational equation as
\begin{equation}
\begin{split}
    \dot{\boldsymbol{\eta}}(t) = \operatorname{D}^{(0)}\textbf{f}\otimes I_N \, \boldsymbol{\eta}(t) -  \operatorname{D}^{(\tau)}\textbf{h}\otimes L \, \boldsymbol{\eta}(t-\tau),
\end{split}
\label{VarEqDiff}
\end{equation}
For a symmetric Laplacian matrix $L$, the system eigenmodes can be directly decomposed into $M$ orthogonal modes \cite{borner2020delay}:
\begin{equation}
\begin{split}
    \dot{\bm\zeta_j}(t) = \operatorname{D}^{(0)} \textbf{f} \,\, \bm\zeta(t) + \nu \operatorname{D}^{(\tau)}\textbf{h} \,\, \bm\zeta_j(t-\tau), 
\end{split}
\label{eq.delayedVarEqDiff}
\end{equation}
where $\nu$ represents the eigenvalues of the Laplacian matrix $L$.

For the corresponding SL model, 
%the dynamical equations are represented in polar form as 
%
% \begin{equation}
% \begin{aligned}
% & \dot{r}_j(t)=\left[\lambda-r_j(t)^2\right] r_j(t)+\sum_{k=1}^M A_{j k}\left[r_k(t-\tau) \cos \left(\phi_k(t-\tau)-\phi_j(t)\right)-r_j(t-tau)\cos \left(\phi_j(t-\tau)-\phi_j(t)\right)\right] \\ 
% & \dot{\phi}_j(t)=\omega-\gamma r_j(t)^2+ \sum_{k=1}^M A_{j k}\left[\frac{r_k(t-\tau)}{r_j(t)} \sin \left(\phi_k(t-\tau)-\phi_j(t)\right)-\frac{r_j(t-\tau)}{r_j(t)} \sin \left(\phi_j(t-\tau)-\phi_j(t)\right)\right].
% \end{aligned}
% \end{equation}
%
we also have that the identical synchronous solution is given by $z(t) = \sqrt{\lambda}\,e^{i\left(\omega-\gamma\lambda\right)t}$, yielding the matrices
\begin{equation}
\begin{aligned}
\operatorname{D}^{(0)}\textbf{f}  = \begin{bmatrix}
-2\lambda & 0\\
-2\gamma \sqrt{\lambda} & 0
\end{bmatrix}, \,\,\,
\operatorname{D}^{(0)}\textbf{h} = 
0_2, \,\,\, \text{and} \,\,\,
\operatorname{D}^{(0)}\textbf{h} = 
\begin{bmatrix}
\cos{(\phi)} & -\sqrt{\lambda} \sin{(\phi)}\\
\frac{1}{\sqrt{\lambda}} \sin{(\phi)} & \cos{(\phi)}\\
\end{bmatrix}.
\end{aligned}
\end{equation}
\noindent 
with $\phi = -\left(\omega-\lambda\gamma\right)\tau$. Thus, the stability of each mode is characterized by the characteristic equation 
\begin{equation}
    \det\left(\operatorname{D}^{(0)}\textbf{f} - \nu \operatorname{D}^{(0)}\textbf{h} \, e^{-\lambda_\ell\tau} - \lambda_\ell I_2\right) = 0. 
\end{equation}
To numerically compute the spectrum $\lambda_\ell$ and the corresponding $\lambda_{\rm max}$, we employ the MATLAB function \texttt{ddebiftool\_stst\_stabil} provided in the DDE-BIFTOOL package for the analysis of delay differential equations \cite{engelborghs2000numerical}. Unlike the MSF analysis proposed in the main text, the MSF analysis of the delay-coupled system \eqref{DynEqDelayDiff} is more straightforward as it does not depend on a choice of in-degree $d^\star$ due to the diffusive nature of the coupling. Figure~\ref{fig:CouplingCompare}(a) shows the MSF landscapes for varying time delays, demonstrating that the stability region tends to shrink as $\tau$ increases. Crucially, the deepest point of the MSF (that is, $\lambda_{\rm max}^* = \argmin_{\nu} \lambda_{\rm max}(\nu)$) is always located in the positive real axis, independently of $\tau$. These results highlight that, for this class of diffusive coupling scheme, the all-to-all network is an optimal solution.

%==================================
\bigskip\noindent 
\textbf{Delayed weakly diffusive coupling.} \rev{We now consider the delay-coupled system \eqref{Dyn_LK_Eq} with ``weakly diffusive coupling,'' such as the form $\textbf{h}(\textbf{x}_j(t),\textbf{x}_k(t-\tau)) = \textbf{h}'(\textbf{x}_k(t-\tau)-\textbf{x}_j(t))$.} We call this coupling scheme weakly diffusive since it vanishes under any of the following conditions: 1) in the absence of delays ($\tau = 0$); 2) the synchronized solution $\mathbf{x}(t) = \mathbf{x}^*$ is an equilibrium point; and 3) the synchronized periodic solution $\mathbf{x}^*(t)$ has period $kT = \tau$, for $k\in\mathbb{N}$. In this case, we can apply the methodology described in the main text to obtain the variational equation \eqref{eq.msf.decomposed}. 
For the SL model \eqref{eq:Dyn_SL}, we obtain the following matrices: 
\begin{equation}
\begin{aligned}
\operatorname{D}^{(0)}\textbf{f} = \begin{bmatrix}
\lambda-3r^{*2} & 0\\
-2\gamma r^* & 0\\
\end{bmatrix},
\,\,
\operatorname{D}^{(0)}\textbf{h} = 
\begin{bmatrix}
-1 & r^* \sin{(\phi)}\\
-\frac{1}{r^*} \sin{(\phi)} & -\cos{(\phi)}\\
\end{bmatrix},
\,\, \text{and} \,\,
\operatorname{D}^{(\tau)}\textbf{h} = 
\begin{bmatrix}
\cos{(\phi)} & -r^* \sin{(\phi)}\\
\frac{1}{r^*} \sin{(\phi)} & \cos{(\phi)}\\
\end{bmatrix},
\end{aligned}
\label{eq.sm.SLDfDh}
\end{equation}
with $\phi=-\Omega_0\tau$. Here, the MSF analysis depends on the choice of minimal indegree $d^\star$. Figure~\ref{fig:CouplingCompare}(b), (c) shows that the stability regions enlarge (shrink) for increasing values of $d^\star$ ($\tau$). Consistent with our theoretical analysis for small $\tau$ in Sec.~\ref{sm:ConditionsForTheory}, these results confirm that $\Omega$ depends monotonically on the choice of in-degree.

%==================================
\bigskip\noindent
\textbf{\rev{General delayed coupling.}} \rev{We now consider delay-coupled systems of the form \eqref{Dyn_LK_Eq} for which the coupling function $\textbf{h}\left(\textbf{x}_j(t),\textbf{x}_k(t-\tau)\right)$ does not vanish for all $\tau\geq 0$.} As examples, we consider both the LK model (as considered in the main text in Fig.~\ref{fig:DeepSize_compare}) and a modified version of the SL model with non-diffusive coupling, so that the numerical results can be directly compared with the types of coupling schemes considered above.

The LK model \eqref{eq.lk} can be represented in polar coordinates as follows:
\begin{equation}
\begin{aligned}
    \dot{r}_j(t)&=\frac{1}{2}\left(G_j-\gamma\right) r_j(t)+ \sum_{k=1}^M {A}_{j k} r_k(t-\tau) \cos \left(\phi_k(t-\tau)-\phi_j(t)\right),
    \\
    \dot{\phi}_j(t)&=\frac{\alpha}{2}\left(G_j-\gamma\right)+ \omega+ \sum_{k=1}^M {A}_{j k} \frac{r_k(t-\tau)}{r_j(t)} \sin \left(\phi_k(t-\tau)-\phi_j(t)\right),
    \\
    \dot{N}_j(t)&=J_{0}-\gamma_{n} N_j(t)-G_j r_j^2(t),
\end{aligned}
   \label{LK_eqs_polar}
\end{equation}
where $r_j(t)$ and $\phi_j(t)$ are respectively the amplitude and phase of the electric field $E_j(t)=r_j(t)e^{i\phi_j(t)}$ of laser $j$, and $N_j(t)$ is the corresponding carrier number. We can apply the methodology described in the main text to obtain the variational equation \eqref{eq.msf.decomposed} around the stationary solution $E_j(t)=r^*e^{i\Omega_0 t}$ and $N_j(t)=N^*$, $\forall j$, with matrices given by
\begin{equation}
\begin{aligned}
\operatorname{D}^{(0)}\textbf{f}  &= \begin{bmatrix}
\frac{1}{2} \left(G_j\frac{ 1-s r_j^{*2}}{1+s r_j^{*2}}-\gamma \right) & 0 &  \frac{r^*_j}{2}  \frac{G_j}{N_j-N_0}\\
-\alpha \frac{G_j}{1+s r_j^{*2}} sr^*_j & 0 & \frac{\alpha}{2}\frac{G_j}{N_j-N_0}\\
-2r^*_j \frac{G_j}{1+s r_j^{*2}} & 0 &  -\left(\gamma_n+ r_j^{*2} \frac{G_j}{N_j-N_0}\right) \\
\end{bmatrix},
\\
\operatorname{D}^{(0)}\textbf{h} &= 
\begin{bmatrix}
0 & r^* \sin{(\phi)} & 0\\
-\frac{1}{r^*} \sin{(\phi)} & -\cos{(\phi)} & 0\\
0 & 0 & 0\\
\end{bmatrix}, \,\, \text{and} \,\,
\\
\operatorname{D}^{(\tau)}\textbf{h} &= 
\begin{bmatrix}
\cos{(\phi)} & -r^* \sin{(\phi)}& 0\\
\frac{1}{r^*} \sin{(\phi)} & \cos{(\phi)}& 0\\
0 & 0 & 0\\
\end{bmatrix},
\end{aligned}
\label{eq.sm.LKDfDh}
\end{equation}
with $\phi=-\Omega_0\tau$. The values of the constructive parameters used throughout the paper are: frequency shift $\omega=0$, linewidth enhancement factor $\alpha=5$, gain coefficient $g=1.5\times 10^{-5} \,\text{ns}^{-1}$, gain saturation coefficient $s=10^{-7}$,  cavity loss $\gamma=500 \,\text{ns}^{-1}$, carrier loss rate $\gamma_n=0.5 \,\text{ns}^{-1}$, carrier number at transparency $N_0=1.5\times 10^8$, pump current $J_{0}=g_{p} \gamma_{n}\left(N_{0} + \frac{\gamma}{g}\right)$, and pump gain $g_{p}=2.55$. These values are consistent with experimentally relevant regimes \cite{kozyreff2001dynamics,liu2008coherent,liu2014nonlinear}.

Similar to the coupling scheme modeled in the LK equations, we consider the following modified SL model with the non-diffusive coupling, described in polar coordinates as
\begin{equation}
\begin{aligned}
& \dot{r}_j=\left[\lambda-r_j(t)^2\right] r_j(t)+\sum_{k=1}^M A_{j k}\left[r_k(t-\tau) \cos \left(\phi_k(t-\tau)-\phi_j(t)\right)\right], \\ 
& \dot{\phi}_j=\omega-\gamma r_j(t)^2+ \sum_{k=1}^M A_{j k}\left[\frac{r_k(t-\tau)}{r_j(t)} \sin \left(\phi_k(t-\tau)-\phi_j(t)\right)\right]. 
\end{aligned}
\end{equation}
Here, the identical solution is given by $r^{*2} = \lambda + d^\star \cos{\Omega_0 \tau}$ and $\Omega_0 = \omega-\gamma(\lambda + d^\star \cos{\Omega_0 \tau}) -d^\star \sin{\Omega_0 \tau}$. Then, the corresponding matrices of the variational equation \eqref{eq.msf.decomposed} are given by 
\begin{equation}
\begin{aligned}
\operatorname{D}^{(0)}\textbf{f}  = \begin{bmatrix}
\lambda-3r^{*2} & 0\\
-2\gamma r^* & 0\\
\end{bmatrix},
\,\,
\operatorname{D}^{(0)}\textbf{h} = 
\begin{bmatrix}
0 & r^* \sin{(\phi)}\\
-\frac{1}{r^*} \sin{(\phi)} & -\cos{(\phi)}\\
\end{bmatrix},
\,\, \text{and} \,\,
\operatorname{D}^{(\tau)}\textbf{h} = 
\begin{bmatrix}
\cos{(\phi)} & -r^* \sin{(\phi)}\\
\frac{1}{r^*} \sin{(\phi)} & \cos{(\phi)}\\
\end{bmatrix},
\end{aligned}
\end{equation}
with $\phi=-\Omega_0\tau$.

For this class of coupling schemes, the MSF analysis also depends on both $d^\star$ and $\nu$. Figure~\ref{fig:CouplingCompare}(d--g) shows the stability regions of the MSF for increasing values of $\tau$ and $d^\star$. For all cases of the SL model, increasing the indegree enlarges the stability region, whereas the opposite trend is observed for the LK model.
%These results indicate that the dependence of the depth and area of the stability region on the indegree is strongly influenced by how the state variables are coupled across oscillators.
To understand this difference, note that the LK model has a single state variable (the carrier number $N_j$) that remains uncoupled to all state variables $(r_k,\phi_k,N_k)$ of other oscillators $k\neq j$. This partial decoupling effectively reverses the size dependence of the stability region on the indegree.

\begin{figure*}%[tbhp]
\centering
\includegraphics[width=0.8\linewidth]{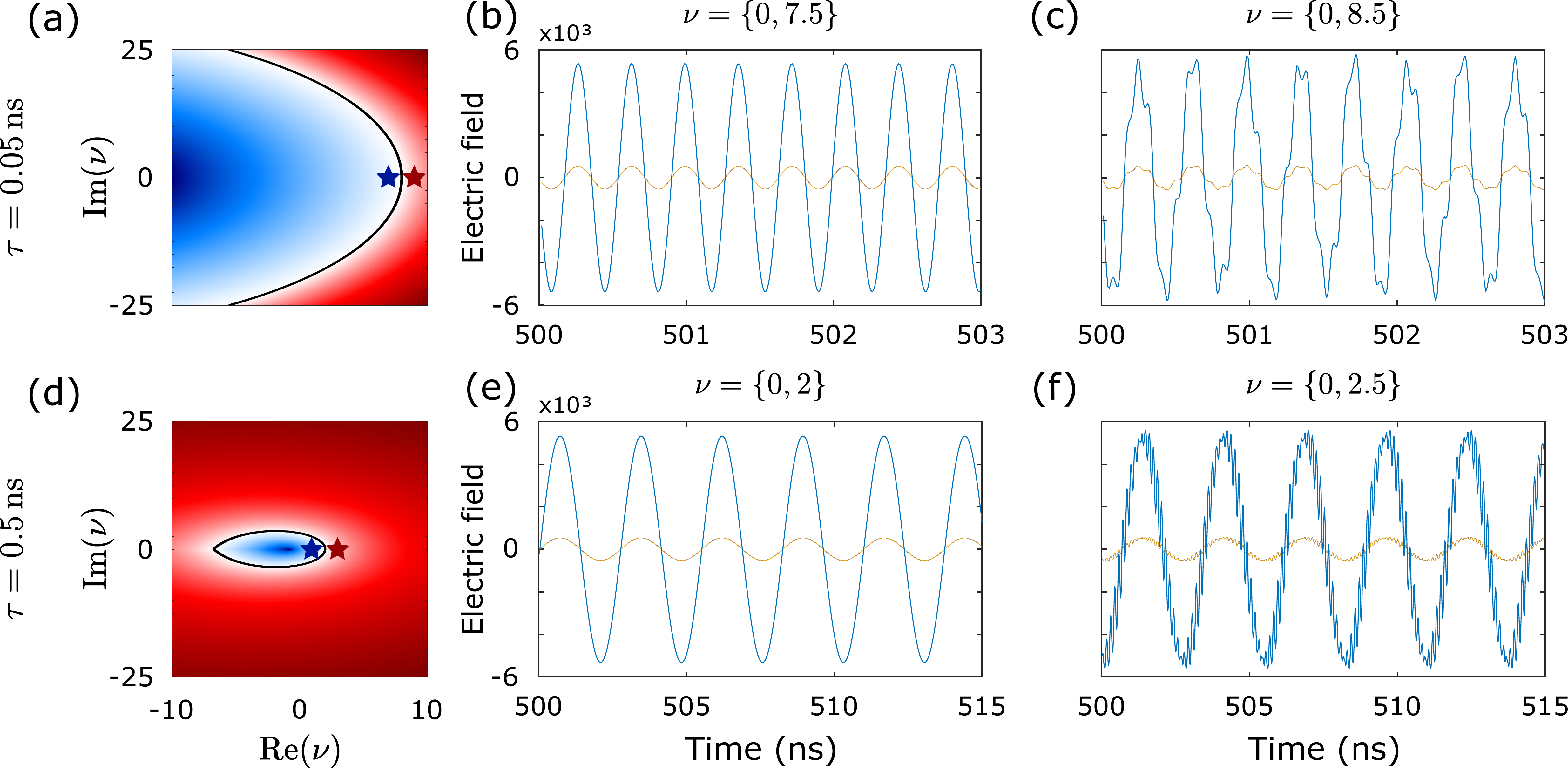}
\caption{\rev{Representative time series of the LK model illustrating stable and unstable synchronization predicted by the MSF. 
(a) MSF landscape for $\tau = 0.05\,\mathrm{ns}$. The blue and red stars indicate the nonzero eigenvalue $\nu$ of an adjacency matrix predicted by the MSF to yield stable and unstable synchronous states, respectively.
(b), (c) Numerical simulations of the LK dynamics for an adjacency matrix with eigenvalues $\nu=\{0,7.5\}$ (b) and $\{0, 8.5\}$ (c). The colored low-amplitude curves illustrate the imaginary parts of the electric fields $E_j$ of the individual oscillators (which largely overlap), whereas the blue curve shows the combined electric field $\sum_j E_j$. Consistent with the MSF prediction in panel (a), the identical synchronous state is stable for $\nu = 7.5$ and unstable for $\nu = 8.5$.
(d)--(f) Same as in panels (a)--(c) for $\tau = 0.5$ ns.}
\label{fig:CouplingCompare_TimeSeries}}
\end{figure*}

%==================================
\bigskip\noindent
\rev{\textbf{Numerical validation of MSF predictions.}
To validate the MSF predictions, Fig. \ref{fig:CouplingCompare_TimeSeries} presents the time-series simulations of the LK dynamics for both small ($\tau=0.05$ ns) and large ($\tau=0.5$ ns) delays.
For $\tau=0.05$ ns, the stability region is relatively broad, allowing stable synchronization over a wide range of eigenvalues $\nu$. As an example, we consider an all-to-all network with $M=10$ nodes and coupling strength $A_{jk} = \kappa$, $\forall j,k$. In this case, the network spectrum is $\nu=\{0,M\kappa\}$. For $\kappa = 0.75$, the spectrum lies entirely within the stability region of the MSF landscape, and, as predicted by the theory, the system converges to an identical synchronous state characterized by a constant oscillation frequency. Increasing the coupling strength to $\kappa=0.85$ shifts the spectrum to $\nu=\{0,8.5\}$, placing it outside the stability region. Again in agreement with theory, synchronization is now unstable, and the electric field converges instead to a multimodal periodic orbit with strong frequency fluctuations.
For large delay ($\tau=0.5$ ns), the stability region becomes substantially narrower, restricting the range of coupling strengths that support synchronization.}

%==================================
\rev{\section{Comparison between delayed and non-delayed MSF}}

\begin{figure}[b]
\centering
\includegraphics[width=0.75\linewidth]{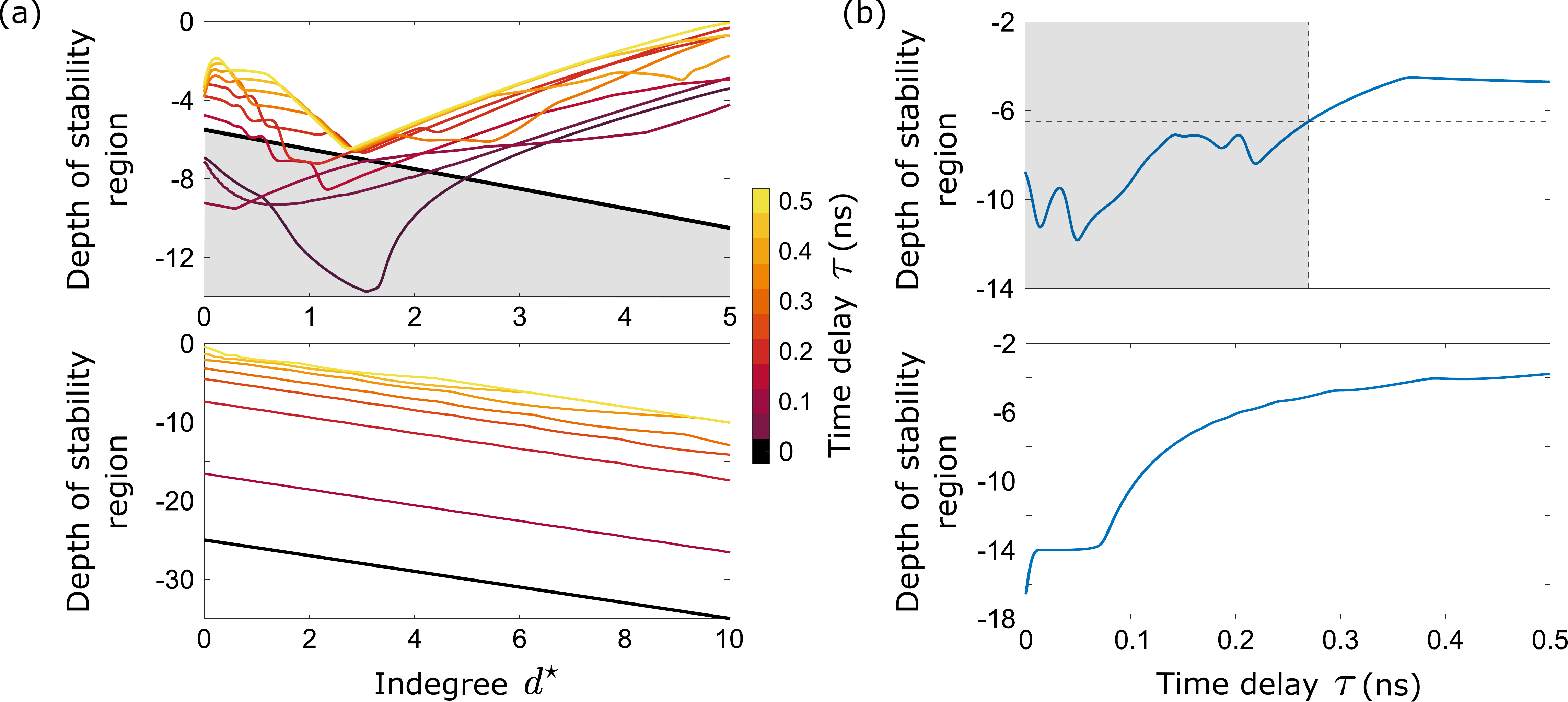}
\caption{\rev{MSF analysis for delayed and non-delayed coupling. (a) Depth of the MSF landscape versus indegree $d^\star$ for the LK (top) and SL (bottom) models. The black curves correspond to $\tau = 0$, whereas the colored curves represent increasing time delays in the range $0.05 \leq \tau \leq 0.5$~ns. (b) Depth of the MSF landscape versus time delay $\tau$ for $d^\star = 1$, again for the LK (top) and SL (bottom) models. The vertical dashed line marks the critical delay below which the delayed system is more stable than its non-delayed counterpart, indicated by the horizontal dashed line. In the top panels, the shaded region highlights the parameter range for which delayed coupling yields greater stability than instantaneous coupling.}
\label{fig:LEtau0}}
\end{figure}

Figure~\ref{fig:LEtau0} further shows that, in the LK model \eqref{eq.lk}, time delays qualitatively alter the MSF landscape. In particular, the effect of the indegree $d^\star$ depends on whether delays are present. For $\tau =0 $, increasing indegree enhances stability and deepens the MSF landscape, while a non-monotonic dependence is observed for $\tau>0$. Moreover, for delays satisfying $\tau \leq 0.27$~ns, there exists a range of $d^\star$ for which maximal stability, corresponding to the deepest point of the MSF landscape, exceeds that of the non-delayed case. This is reflected in the intervals where the delayed (colored) curves lie below the non-delayed (black) curve. \rev{For the SL model \eqref{eq:Dyn_SL}, increasing the indegree monotonically improves stability for all choices of $\tau$ (which is in agreement with our analytical results in Sec.~\ref{sm:ConditionsForTheory} for the case of small $\tau$). However, in contrast to the LK model, the delayed SL models never outperform the non-delayed case: increasing the time delay consistently reduces the depth of the MSF landscape, leading to a progressive loss of stability.}

%==================================
\bigskip \bigskip
\section{Network optimization method}

\label{sm:NetOpt}
Algorithm~\ref{alg.coupling_opt} introduces an iterative procedure to tune the edge weights of the adjacency matrix $A$ in order to maximize the stability of the synchronous solution $\mathbf{x}^*$, which we then denote as ${\mathbf x}^{(0)}$ at the initial step. The algorithm begins with an initial guess $A^{(0)}$ and iteratively refines $A$ through small, constrained perturbations $\delta A$ that minimize $\lambda_{\rm max}$ as formulated in Eq.~\eqref{eq.optimizaitonfunction}. In each iteration step $k$, we find the stationary solution ${\mathbf x}^{(k)}$ by solving an implicit function $\textbf{G}( \textbf{x}^{(k)},A^{(k)})=0$; using the LK model as an example, the synchronous state is obtained by solving a set of transcendental equations derived by substituting $(\dot r_j,\dot \phi_j, \dot N_j) = (0,\Omega_0,0)$ into Eq.~\eqref{LK_eqs_polar}. The scaling factor $\gamma$ controls the magnitude of these perturbations, which should be sufficiently small to ensure that the numerical continuation of $\textbf{x}^*$ remains within a specified tolerance $\epsilon_1$; this step also ensures that the obtained stationary solutions $\mathbf{x}_j^*$ are nearly identical across all oscillators. Additional constraints to the network structure can be incorporated into the optimization problem \eqref{eq.optimizaitonfunction}, such as enforcing $\sum_{i,j}\delta A_{ij} = 0$ (Fig.~\ref{fig:Networks}, green curve) or imposing that $\delta A_{ij} = 0$ if $A_{ij}^{(0)} = 0$ (Fig.~\ref{fig:Networks}, purple and cyan curve). To obtain the results presented in Fig. \ref{fig:Networks}, we set the following hyperparameters: $\gamma = 1$, $T = 100$, $\epsilon_1 = 1$, and $\epsilon_2 = 10^{-4}$. For each of the four optimization scenarios, Algorithm~\ref{alg.coupling_opt} was run over 100 independent realizations. In Fig.~\ref{fig:Networks}(a), we report the optimization steps corresponding to the realization that achieved the smallest MTLE.

%----------------------------------------------------------------------------------

\begin{algorithm}[H]
\caption{Network optimization method}
\begin{algorithmic}[1]
\State \textbf{Input:} Initial adjacency matrix \( A^{(0)} \)
\State \textbf{Output:} Optimal adjacency matrix \( A^{(k)} \)
\State Determine the identical synchronous solution \( \mathbf{x}^{(0)} \) by solving the implicit function 
\begin{equation}
\textbf{G}( \textbf{x}^{(0)},A^{(0)})=0
\end{equation}
\State Compute the MTLE \(\lambda_{\rm max}^{(0)}(\textbf{x}^{(0)},A^{(0)}) \) by solving the characteristic equation \eqref{CharacteristicEq} for \(\textbf{x}^{(0)}\) and \(A^{(0)}\)
\For{$k = 1$ to $T$ (\# trials)}
    \State Generate a perturbation vector $\delta \bar A$ containing all non-diagonal elements \( \delta A_{ij} \sim \mathcal N(0,1), \forall i\neq j \)
    \State Normalize the perturbation vector such that $\|{\delta \bar A}\| = \gamma$, where $\gamma$ is a scaling factor
    \State Solve the optimization function
 \begin{equation}
     \begin{aligned}
     \delta A^{(k)} = &\argmin_{\delta A} \lambda_{\rm max}\left(\textbf{x}^{(k)},A^{(k-1)}+\delta A\right) 
     \\
     & \,\, \text{s.t.} \,\,\, \tilde{\textbf{G}}\left(\textbf{x}^{(k)},A^{(k-1)}+\delta A\right) = 0, \,\,\,  A_{ij}^{(k-1)}+\delta A_{ij}\geq 0, \forall (i,j), \,\,\, \|\delta A\| = \gamma
     \end{aligned}
     \label{eq.optimizaitonfunction}
 \end{equation}
    \While{$\| \mathbf{x}^{(k)} - \mathbf{x}^{(k-1)} \| \geq \epsilon_1$}
        \State Reduce the scaling factor $\gamma \leftarrow 0.8\gamma$
        \State Re-solve optimization
    \EndWhile
    \State Update \( A^{(k)} \leftarrow A^{(k-1)} + \delta A \)
    \State Update \( \lambda^{(k)}_{\rm max}(\textbf{x}^{(k)},A^{(k)}) \)
    \If{$|\lambda_{\text{max}}^{(k)} - \lambda_{\rm max}^{(k-1)}| \leq \epsilon_2$}
        \State \textbf{break}
    \EndIf
\EndFor
\end{algorithmic}
\label{alg.coupling_opt}
\end{algorithm}

\begin{figure}[h]
\centering
\includegraphics[width=0.8\linewidth]{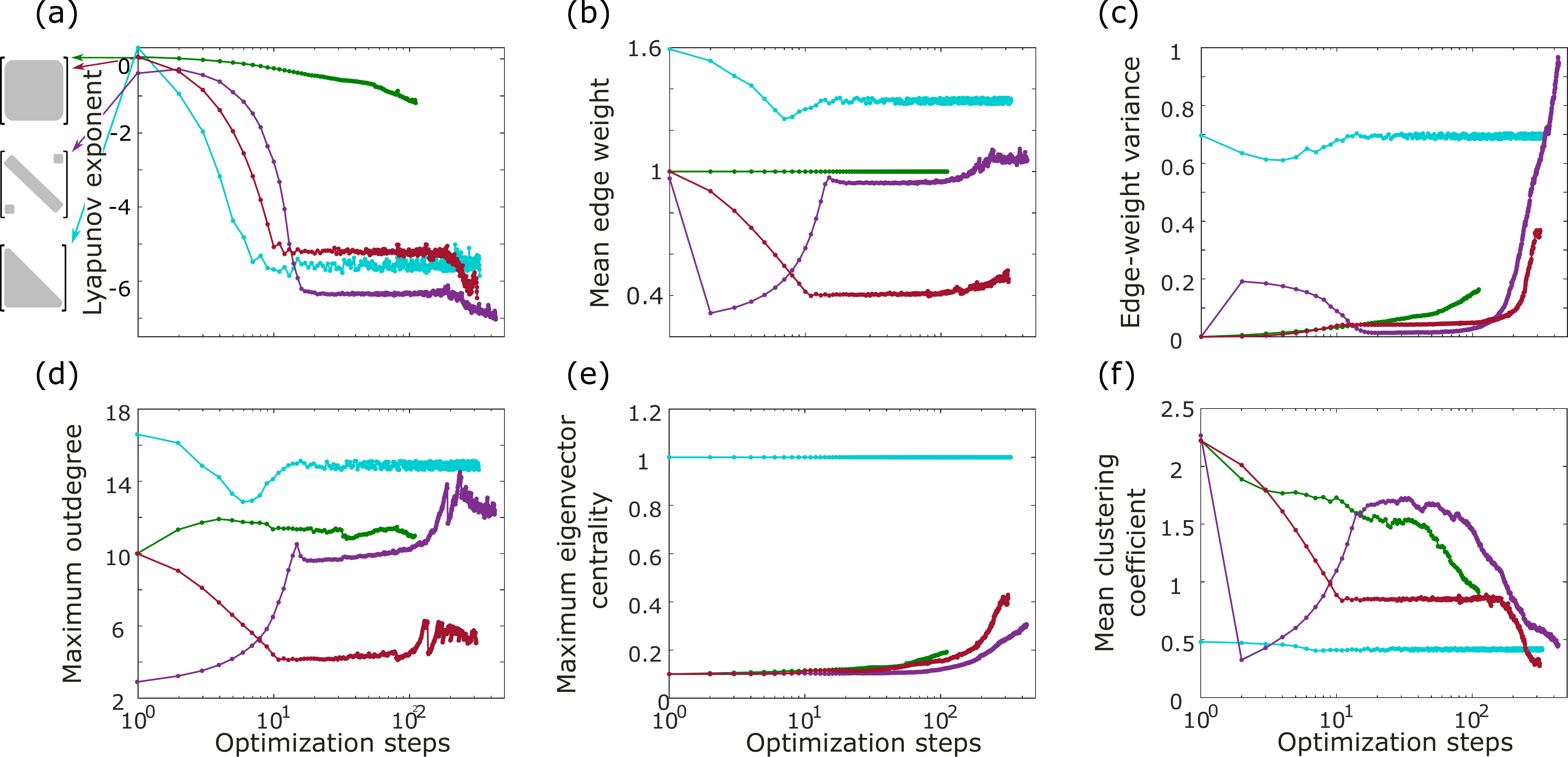}
\caption{\rev{Structural properties during optimization for the LK model.
(a)~MTLE for different initial network configurations and optimization constraints, as in Fig.~\ref{fig:Networks}.
(b--f) Evolution of different network characteristics throughout the optimization process: mean edge weight (b), edge-weight variance (c), maximum outdegree (d), maximum eigenvector centrality (e), and mean clustering coefficient (f).}
\label{fig:structural_indicators}}
\vspace{-0.3cm}
\end{figure}

%==================================
\noindent
\textbf{\rev{Evolution of the network structure during the optimization process.}}
\rev{Figure \ref{fig:structural_indicators} provides further insight into how the network structure evolves throughout the optimization process shown in Fig.~\ref{fig:Networks}. It shows that the mean edge weight is comparable throughtout the optimization steps for most strategies [Fig.~\ref{fig:structural_indicators}(b)], indicating that the improvement in stability is not driven by an overall increase or decrease in coupling strength.  The first optimization step of the ring topology (purple curve) is a notable exception, as the addition of a single edge substantially changes the mean edge weight as well as the edge-weight variance and mean clustering coefficient. In contrast with the mean edge weight, its variance tends to increase sharply during the optimization [Fig.~\ref{fig:structural_indicators}(c)], revealing a progressively more heterogeneous distribution of coupling strengths. This trend is typically accompanied by an increase in the maximum outdegree relative to the mean outdegree [Fig.~\ref{fig:structural_indicators}(d)], consistent with the emergence of nodes that play a highly influential role in the network, and an increase in the maximum eigenvector centrality [Fig.~\ref{fig:structural_indicators}(e)], indicating that a small subset of nodes becomes increasingly influential. Moreover, the mean clustering coefficient tends to decrease throughout the optimization [Fig.~\ref{fig:structural_indicators}(f)], suggesting the formation of more hierarchical architectures. Taken together, these structural changes reveal that optimization drives the network toward a heterogeneous, directional, and hierarchical organization.}

\begin{figure}[b!]
\centering
\includegraphics[width=0.6\linewidth]{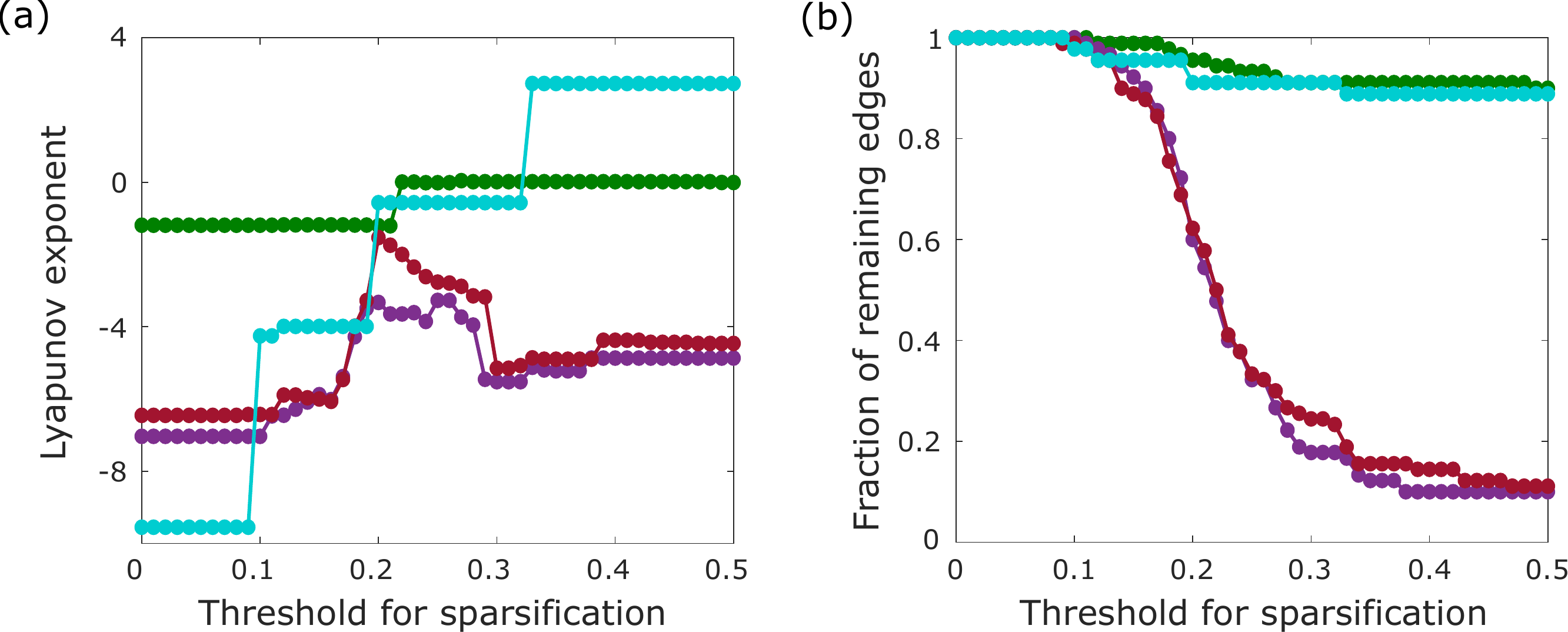}
\caption{Sparsification of the optimal network configuration.
(a) Lyapunov exponent $\lambda_{\rm max}$ and (b) fraction of remaining edges, plotted as functions of the sparsification threshold $\epsilon$. The results are shown for the four optimal networks obtained in Fig.~\ref{fig:Networks}(b): constrained weights, all-to-all initially (green); unconstrained weights, all-to-all initially (red); unconstrained weights, ring initially (purple); and existing edges only, directed tree-like (cyan).
}
\label{fig:NetworksSparse}
\vspace{0pt}
\end{figure}

%==================================
\bigskip\noindent
\textbf{Sparsification of the optimal network structures.}
Motivated by the dominance of a few large-weight edges in the optimal networks obtained in Fig.~\ref{fig:Networks}(b), we investigate the impact of \textit{sparsification} on synchronization stability. Specifically, we remove all edges with weights below a threshold $A_{ij}\leq \epsilon$ and rescale the remaining edges to preserve the original total weight.
Figure \ref{fig:NetworksSparse} shows the dependence of the Lyapunov exponent $\lambda_{\rm max}$ on the sparsification threshold $\epsilon$. In all cases, increasing $\epsilon$ degrades stability, highlighting the critical role of low-weight edges in promoting synchronizability. However, the sensitivity to sparsification depends strongly on the network classes. In the constrained all-to-all case, $\lambda_{\rm max}$ remains nearly invariant to $\epsilon$, as only a few edges are trimmed due to the system's high overall coupling. For unconstrained optimizations (both all-to-all and ring topologies), stability decreases with increasing sparsification; notwithstanding, we note that $\lambda_{\rm max}$ remains below its pre-optimization value even after removing nearly 90\% of the edges. In the master-slave configuration, a sharp loss of stability is observed: pruning even a small fraction of edges induces an abrupt transition to instability. %Together, these results suggest a complex interplay between synchronizability and structure in directed systems.

\end{document}